\documentclass[a4paper,11pt,twoside]{article}

\usepackage[T1]{fontenc}
\usepackage{amssymb}
\usepackage{AlegreyaSans}
\usepackage{lmodern}
\usepackage{palatino}
\usepackage{alltt}

\usepackage{lscape}
\usepackage{section}
\usepackage{headerfooter}
\usepackage{table}
\usepackage{justify}
\usepackage{toc}
\usepackage{list}
\usepackage{arc-headerfooter}

\newcommand{\draftfooter}{}
\usepackage{arcemptypage}

\newcommand{\theauthors}{A.\ Keller, C.\ W{\"a}chter, M.\ Raab, D.\ Seibert, D.\ van Antwerpen, J.\ Kornd{\"o}rfer, and L.\ Kettner}

\newcommand{\documenttitle}{The Iray Light Transport Simulation and Rendering System}
\newcommand{\maintitle}{\documenttitle}
\newcommand{\subtitle}{\theauthors\\
NVIDIA\\
\small e-mail: akeller$|$cwaechter$|$mraab$|$dseibert$|$dietgerv$|$jkorndoerfer$|$lkettner@nvidia.com}

\newcommand{\docversion}{Document version 1.0} %
\newcommand{\docdate}{May 3, 2017} %
\newcommand{\copyrightfooter}{}
\let\oldsection\section
\renewcommand{\section}{\cleardoublepage\thispagestyle{arcnoheader}\vspace*{0.5in}\oldsection}

\usepackage[%
    margin=5mm,%
    textfont={small},%
    labelfont={small,bf,rm}%
]{caption}

\usepackage{float}
\usepackage[pdftex]{graphicx}
\usepackage{overpic}
\usepackage{xcolor}
\usepackage{microtype}

\usepackage[
backend=biber,
style=alphabetic,
citestyle=alphabetic,
doi=false,
isbn=false,
url=false,
]{biblatex}

\bibliography{iray,Alex}

\usepackage[hidelinks,pdfpagelayout=TwoPageRight]{hyperref}
\hypersetup{
	colorlinks=false%
	,linkcolor=blue%
	,citecolor=blue%
	,filecolor=blue%
	,urlcolor=blue%
	,pdftitle=\documenttitle
	,pdfauthor=\theauthors
	,pdfcreator=NVIDIA ARC GmbH
	,pdfproducer={NVIDIA Klammertext}
	,pdfsubject=%
	,pdfkeywords=%
	,plainpages=false}

\pdftrailerid{}
\begin{document}
\pagenumbering{roman}\setcounter{page}{1}

{\sfont{20pt}\maintitle}\\[10pt]
{\sfont{13pt}\subtitle}\\[7mm]
\docdate \\
\docversion
\vfill
\begin{minipage}{\linewidth}\centering%
\includegraphics[width=\linewidth]{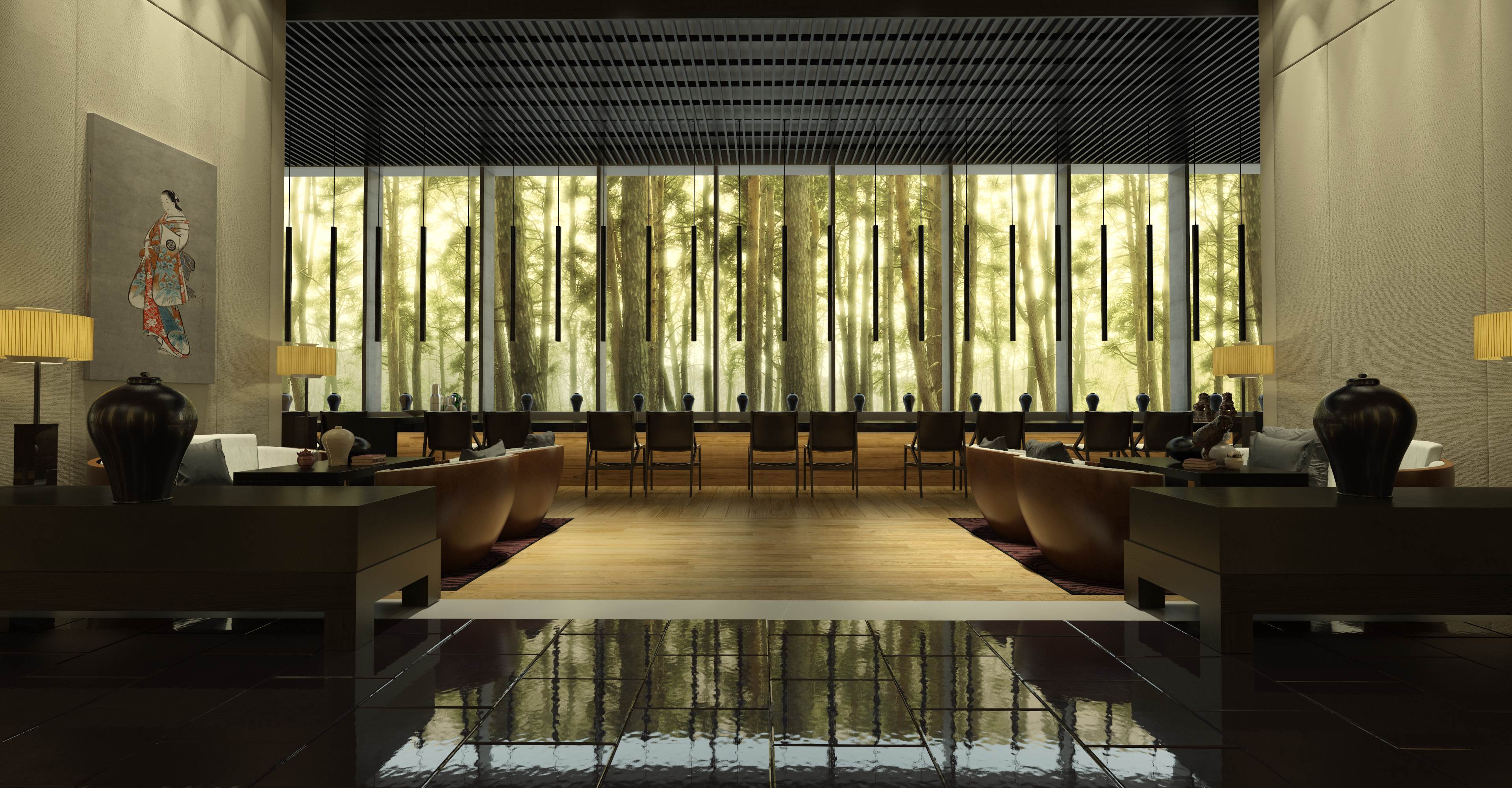}
\end{minipage}

\hfill{\it\scriptsize }
\vfill

\vfill

\hfill\includegraphics[width=200px]{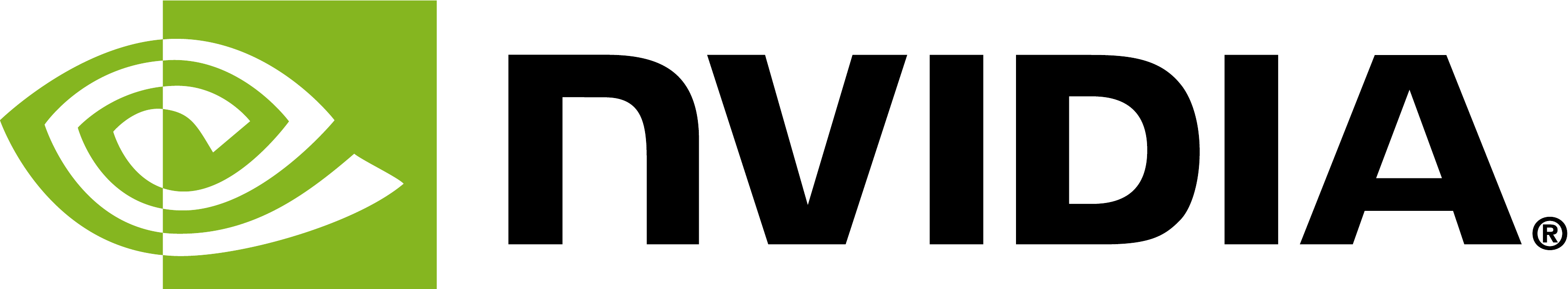}

\hspace*{-10pt} \vspace*{-4pt} {\color{white}NVIDIA}

\clearpage
\normalsize

\newpage
\pagenumbering{roman}\setcounter{page}{2}

\thispagestyle{arcnoheader}
\leavevmode\vfill
\vbox{}
\vfill
\clearpage 
\pagenumbering{roman}\setcounter{page}{3}
\pagestyle{arcnoheader}
\begin{spacing}{0.85}
\renewcommand{\baselinestretch}{0.8}
\tableofcontents
\end{spacing}
\pagestyle{arcpagestyle}

\pagenumbering{arabic}\setcounter{page}{1}

\date{2017}

\section*{Abstract}

While ray tracing has become increasingly common
and path tracing is well understood by now, a major challenge lies in
crafting an easy-to-use and efficient system implementing these technologies.

Following a purely physically-based paradigm while still allowing for
artistic workflows, the
Iray light transport simulation and rendering system \cite{irayrender}
allows for rendering complex
scenes by the push of a button and
thus makes accurate light transport simulation widely available.
In this document we discuss the challenges and implementation choices that follow from our primary design decisions, demonstrating that such a rendering system
can be made a practical, scalable, and efficient real-world application that
has been adopted by various companies across many fields and
is in use by many industry professionals today.

\clearpage

\section{Introduction}

The ultimate goal of light transport simulation is to create
images that cannot be distinguished from measurements of
reality, most commonly in the form of photographs. 
This quest started in the early 1980s \cite{CornellBox}, where simulation results
were compared to photographs of the physical Cornell Box.
Later on, the RADIANCE system \cite{War:94} used more advanced algorithms
to close the gap between reality and simulation, and was
successfully verified in many experiments.

\begin{figure}
  \centering
   \includegraphics[width=0.4915\linewidth,trim={0 2cm 0 0},clip]{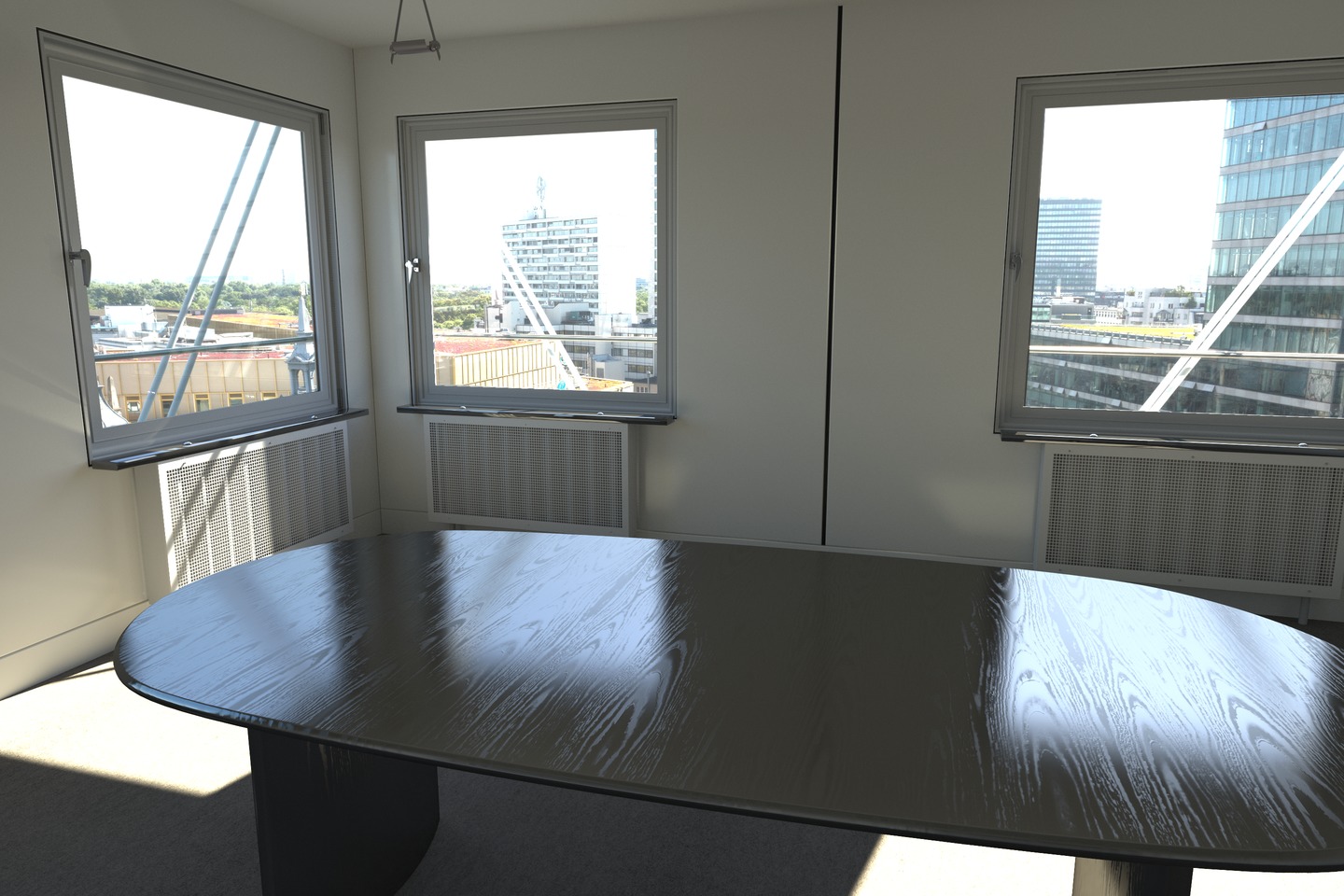} \hfill
   \includegraphics[width=0.4915\linewidth,trim={0 2cm 0 0},clip]{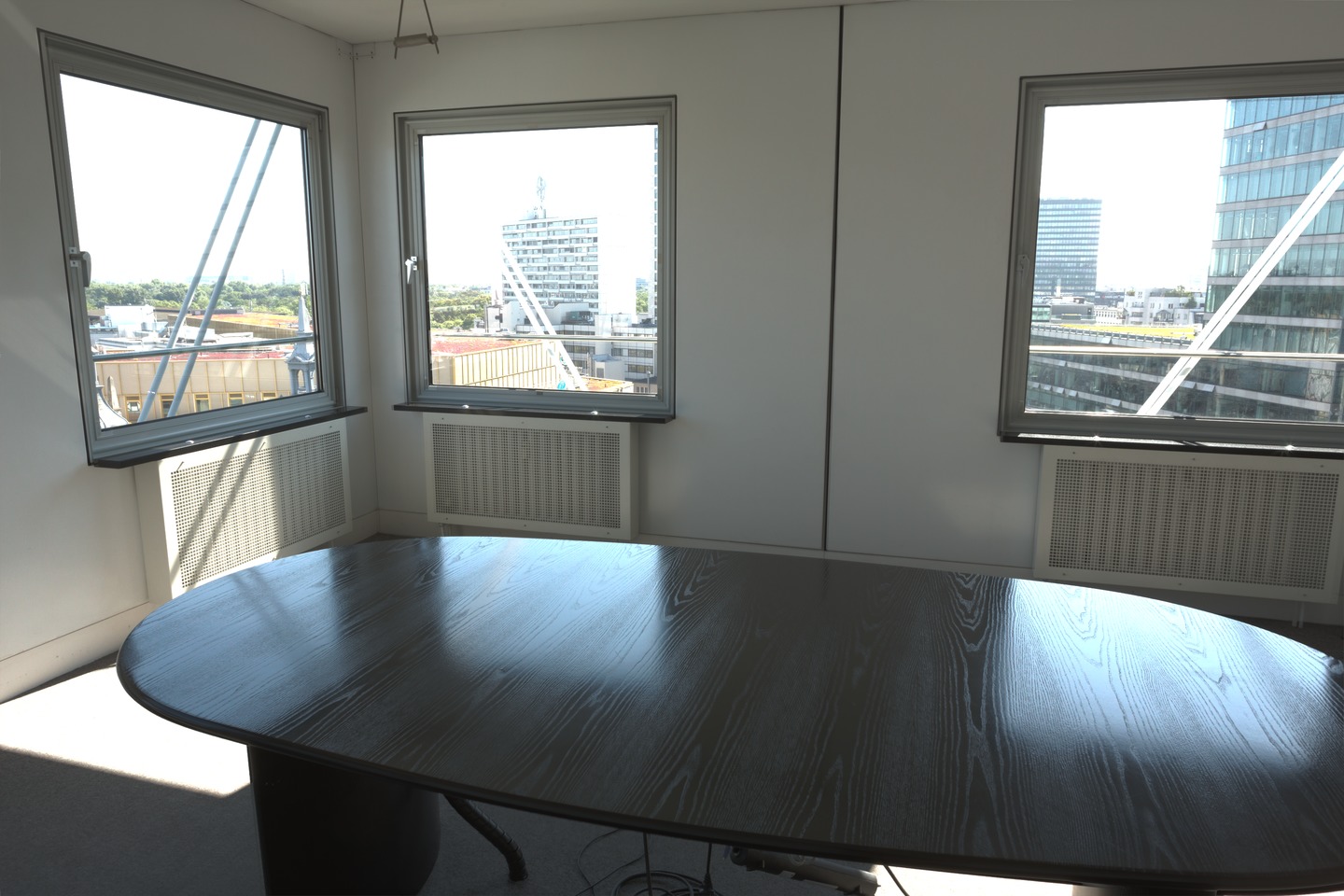}
   \caption{Comparison of a rendered image (left) with a reference photograph (right). %
   The rendering uses a photographic backplate for the visible scenery outside the windows. %
   \label{Fig:RealityCheck}}
\end{figure}

In the same spirit, Fig.~\ref{Fig:RealityCheck} features such a
reality check experiment conducted with Iray, which
is a full-featured light transport simulation
and rendering system that is capable of predictive rendering at the push of a button. While many
contemporary rendering systems can achieve
similar results, they often require extensive parameter tuning
and user expertise to do so.

Obviously, simulation algorithms must be based on the description of
physical entities. However, artistic and non-physical workflows
are a requirement.
The challenge lies in combining these features, which seem to be at odds.

\subsection{Design Decisions} \label{Sec:Design}

With a primary focus on applications in the design industry,
the Iray light transport simulation and rendering system has been
developed over the last decade. Its primary design decisions are:
\begin{description}
\item[Push-button:] As simple as a camera, the push of a button
must deliver the desired image with as few parameters as possible. 
Rendering artifacts must be transient and vanish over time.
Therefore, consistent numerical algorithms (see Sec.~\ref{Sec:Halton})
have been selected as the
mathematical foundation of Iray, as these guarantee the approximation
error to decrease as the sample size increases.

\item[Physically-based rendering:] To generate realistic and predictive
imagery, the rendering core of Iray uses light transport simulation
based on a subset of bidirectional path tracing \cite{VeachPhD}.
Illumination by a large number of geometric light sources with
spatially varying emission, illumination by high-resolution environment maps
with respect to the shading normal, photon aiming, and complex
layered materials have led to the development of new importance sampling techniques.
Details, in particular on the
used importance sampling techniques, are discussed in Sec.~\ref{Sec:Techniques}.

\item[Separating material description from implementation:]
In order to avoid performance pitfalls of user-written
shader code but still grant all freedom for
look development, appearance modeling, and programmable
materials in general, the
material definition language (MDL) (see Sec.~\ref{Sec:MDL})
has been co-designed with the system architecture. By its
strict separation of material description and implementation, efficiency
improvements found in Sec.~\ref{Sec:Techniques} can be implemented within the renderer
and material descriptions can be shared
with other renderers.

\item[Scalability:] The architecture of the system must be such that performance scales across large clusters of
heterogeneous parallel processors (CPUs and GPUs) without sacrificing any of the functionality of a sequential version.
Consistency of the results must be guaranteed even across
changing heterogeneous parallel computing environments including network clusters and
the cloud. Load balancing must be fully automatic and elastic, providing both interactive as well as batch rendering.
This is enabled by deterministic quasi-Monte Carlo methods.
Sec.~\ref{Sec:Parallel} describes Iray's parallel architecture.

\item[Rendering workflows:] The system has
been designed to allow for the natural and seamless combination
of modern artistic and physically-based workflows, including detailed control over outputs that are useful in post-processing via Light Path Expressions (see Sec.~\ref{Sec:Workflows})
and direct, progressive rendering of matte objects without sacrificing the consistency of the light transport simulation.
\end{description}

Many of the design decisions have in common that they imply generalization of some sort, which contrasts with most other renderers. 
For example, the push-button requirement precludes the renderer from employing overly specialized methods
that require manual user input in the form of parameters or choice of the algorithm used. 
Both physically-based rendering and explicit material description fit well to this idea.

While generalized methods are universally desirable, they are often slow to compute, which is why Iray's
performance is integral to enabling this approach in practice.
Thus, the key aspects of the implementation of the design decisions in the Iray light transport simulation
and rendering system are detailed in this document.
Some details that do not directly follow from the objectives outlined above, but are still critical to the implementation of
the practical rendering system, are described in the Appendix.

\section{Light Transport Simulation} \label{Sec:Techniques}

In this section, we discuss the architecture of Iray's light transport simulation, which forms the foundation for the subsystems in the following sections. 

Physically-based rendering, in the form of two-way path tracing
\cite{TwoWayPathTracing} -- a subset of bi-directional path tracing
\cite{LafortunePhD,VeachPhD}, comprising a forward path tracer and a
light tracer -- forms the algorithmic foundation of Iray.
In Iray, path and light tracing are separate stages that are
combined using multiple importance sampling (MIS) \cite{VeachPhD}.
Furthermore, light tracing is completely optional, as not all scenes feature
difficult caustic effects, where path tracing is insufficient in terms of performance and convergence.
In case light tracing is enabled, the same number of paths is simulated in both the light and path tracing step.
Due to this fixed balance, scenes that do not benefit from light tracing will converge faster when disabling it.

The choice of path and light tracing instead of a general bi-directional path tracer is a compromise on code complexity and state storage requirements for massively parallel computation.
To still be able to handle many difficult real world phenomena without
compromising on precision and efficiency, it is key to employ a
set of variance reduction techniques, in particular importance
sampling and next event estimation (NEE), on top of the base algorithm.

The life of a light transport sample in Iray is illustrated in Fig.~\ref{Fig:sampleflowchart}:
The core components are ray tracing, next
event estimation, environment and matte object handling, and material
evaluation and sampling.
Next event estimation in Iray consists of light source importance
sampling and visibility tests -- which can include (partial)
material evaluation to determine transparency.
Note that, while the state chart describes the path tracing step, the light
tracing counterpart is very similar in structure.

The choice of stages is guided by the wavefront architecture, which we will describe in Sec.~\ref{Sec:StateMachine}.
In this section we will focus on the simulation of one light transport path.

Although many components of light transport simulation are relatively straightforward to implement
\cite{PharrHumphreys}, we believe that some specific algorithmic
choices are unique to Iray and worth sharing, in particular
those that do not have exact counterparts in literature.
In the following we will outline some of those implementation aspects
along with further information and rationale.

\subsection{Light Importance Sampling} \label{Sec:LightSources}

\begin{figure}
  \centering
  \includegraphics[height=0.2135\linewidth]{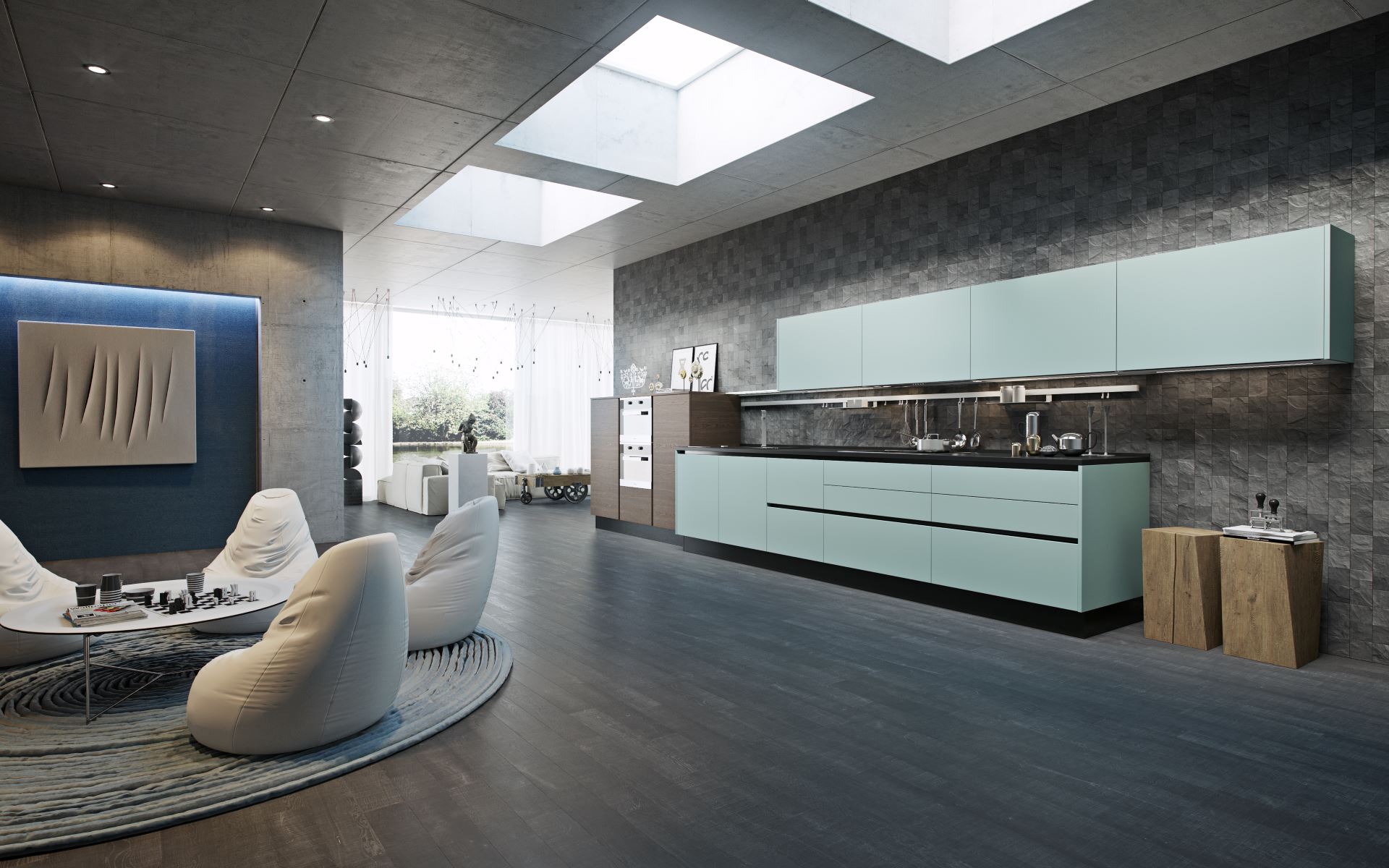} \hfill
  \includegraphics[height=0.2135\linewidth]{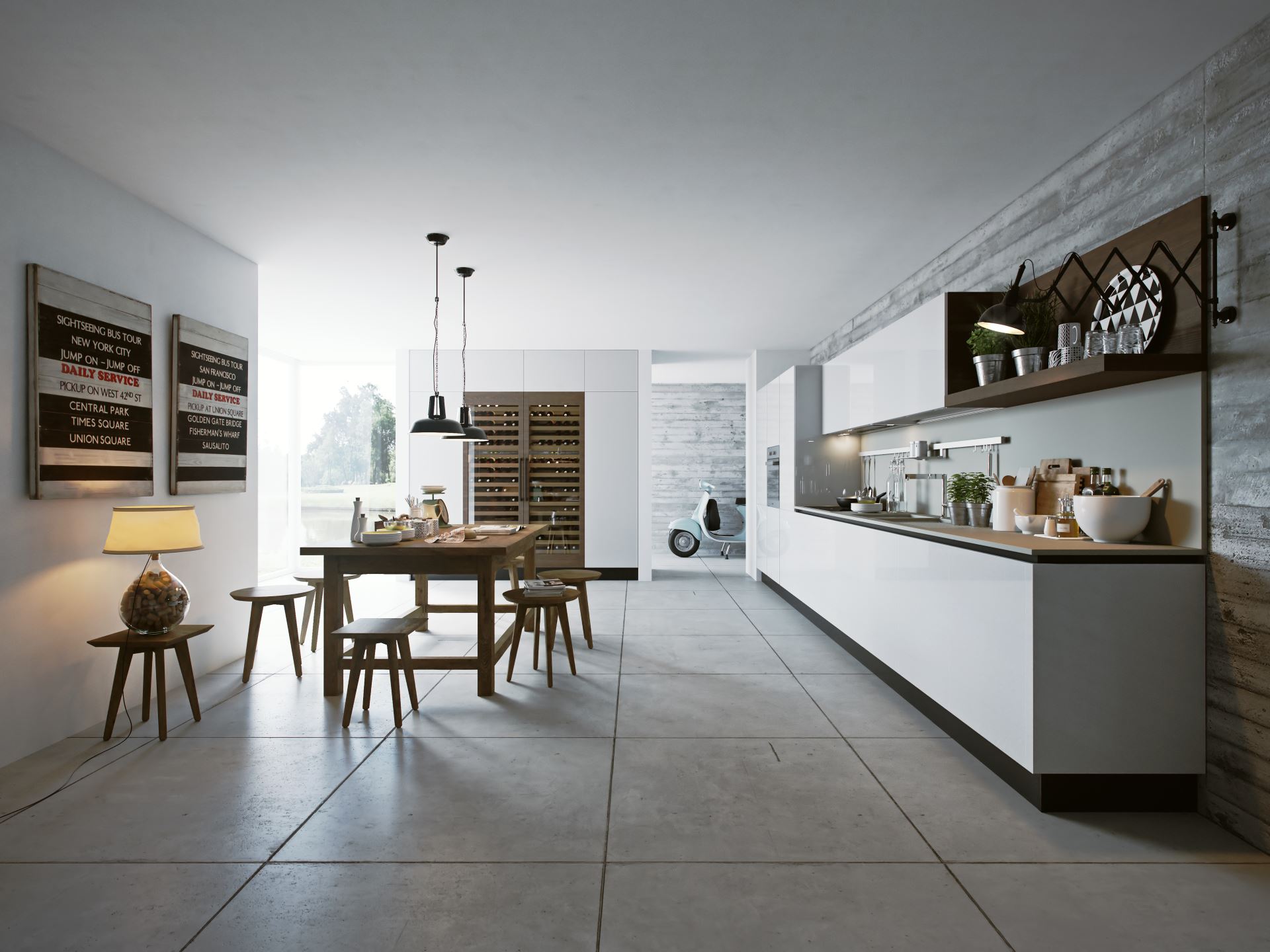} \hfill
  \includegraphics[height=0.2135\linewidth]{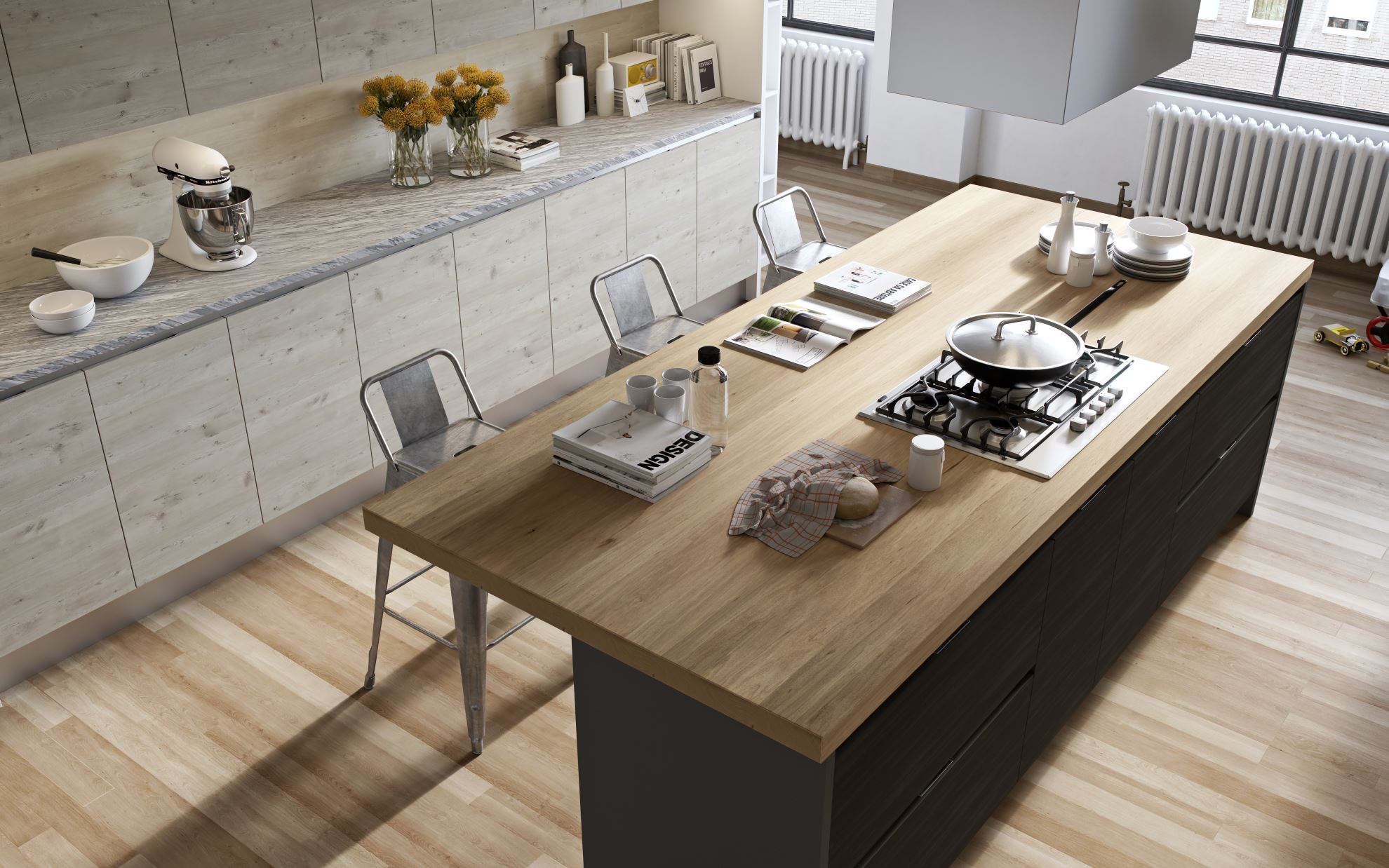}
  \caption{Interiors lit by both the environment and geometric light sources. Images courtesy of Delta Tracing. \label{Fig:DeltaTracing}}
\end{figure}

Importance sampling of all sources of light is a requirement for an efficient path tracer.
This is especially important for design and architecture visualization where scenes often contain many area light sources in conjunction with a measured sky dome (Fig.~\ref{Fig:DeltaTracing}).
Our methods to accelerate importance sampling for geometric lights as well as environment lights have in common that they both rely on specialized hierarchical acceleration data structures.
Those hierarchies are focused specifically on efficient importance sampling, as opposed to hierarchical approaches targeted only at speeding up the evaluation from many sources \cite{WFABDG05}.
We will now discuss both methods.

\subsubsection{Geometric Light Sources}

For area light sources we based the design of our algorithm on the following requirements:
\begin{itemize}
\item
  The number of light sources may be large (tens of thousands).
\item
  The area shape of emitting geometry may be complex, i.e.\ arbitrary meshes.
\item
  Intensities of light sources may vary spatially, especially, when driven by an MDL function.
\end{itemize}
As a consequence we decided against having specialized sampling strategies for a set of uniformly emitting surface shapes, like spheres or rectangles \cite{Shirley:1996:MCTDL}, but rather work with triangles exclusively.
Assuming that the tessellation of the geometry is generally fine enough, spatial variation can implicitly be handled efficiently by assigning a single flux value per triangle as obtained from integrating the intensity function over a triangle's area in a preprocess.
Effectively, this means that we end up with a (potentially large) set of triangle light sources that we need to handle.

The algorithm we employ utilizes a bounding volume hierarchy that is traversed probabilistically. 
Practical scenes usually feature rather small and well separated light sources, such that the bounding volume hierarchy build can quickly encapsulate small regions, wrapping single light sources.

\begin{figure}
  \centering
  \includegraphics[width=\linewidth]{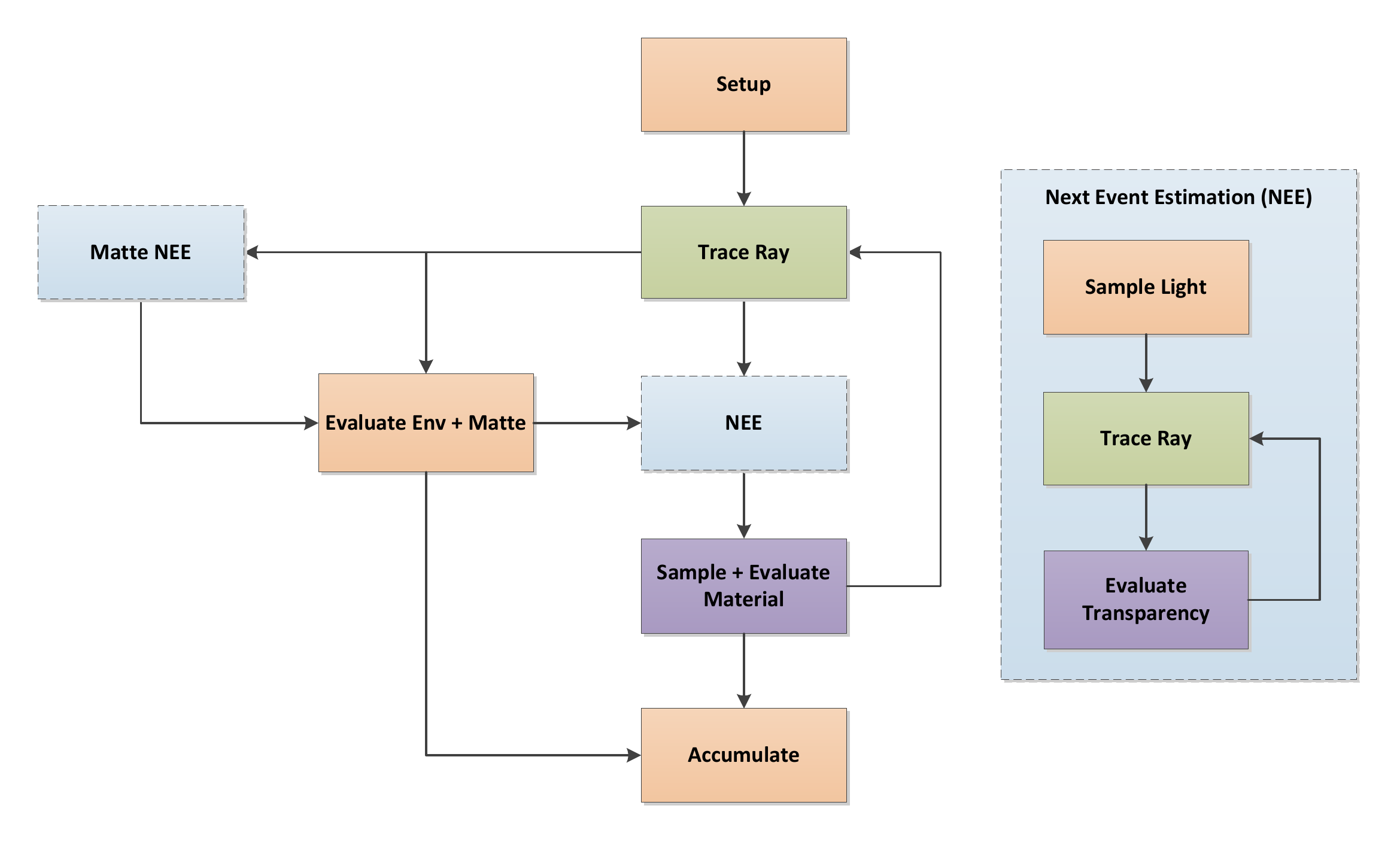}
  \caption{State chart for sample generation. \label{Fig:sampleflowchart}}
\end{figure}

Its fundamental concept is best illustrated by looking at one traversal step:
Given a shading point and the current node in the hierarchy, we estimate the contributions of both child volumes of emitting geometry for that point.
The estimates then drive the probability of which either volume is to be selected to continue the traversal.
Once a leaf containing few triangles is encountered, we importance sample from them according to their estimated flux.
In addition, the hierarchy can be traversed deterministically when a light source is hit in order to compute the probability density of sampling that point for MIS.
For that we utilize the unique traversal order of the leaf and use the previous surface point and normal of the currently simulated transport path to compute the estimates.

In the ideal case, the algorithm for estimating the contribution from a bounding volume of emitting geometry would only need to deal with a single small and far away triangle, where it could easily approximate its contribution using the flux and geometric term, i.e.\ the cosines on surface and light source multiplied by the inverse squared distance.
As a generalization for a volume containing many scattered triangles, we use the combined flux of all contained triangles, the distance to the bounding box center, and a conservative estimate for the cosine on the surface point.
Further, we clamp probabilities to avoid over- or undersampling due to suboptimal estimates and to prevent any bias that could be introduced by probabilities equal to zero.
While for the upper levels of the hierarchy the resulting estimates are relatively rough (but typically still better than random selection), the lower levels quickly yield very good selection probabilities.

To account for directional emission of a bounding volume hierarchy node, as caused by the cosine on the light surface, surfaces that do not emit on the back side, and directional spot-lights and IES emission distribution functions in MDL, we incorporate some additional information into the acceleration data structure:
Instead of having just a single flux value per node, we subdivide the unit sphere into a small set of regions and store one representative value per region.
Those values are then utilized in the traversal to obtain more accurate estimates.

\begin{figure}
	\centering
	\includegraphics[width=0.4915\linewidth]{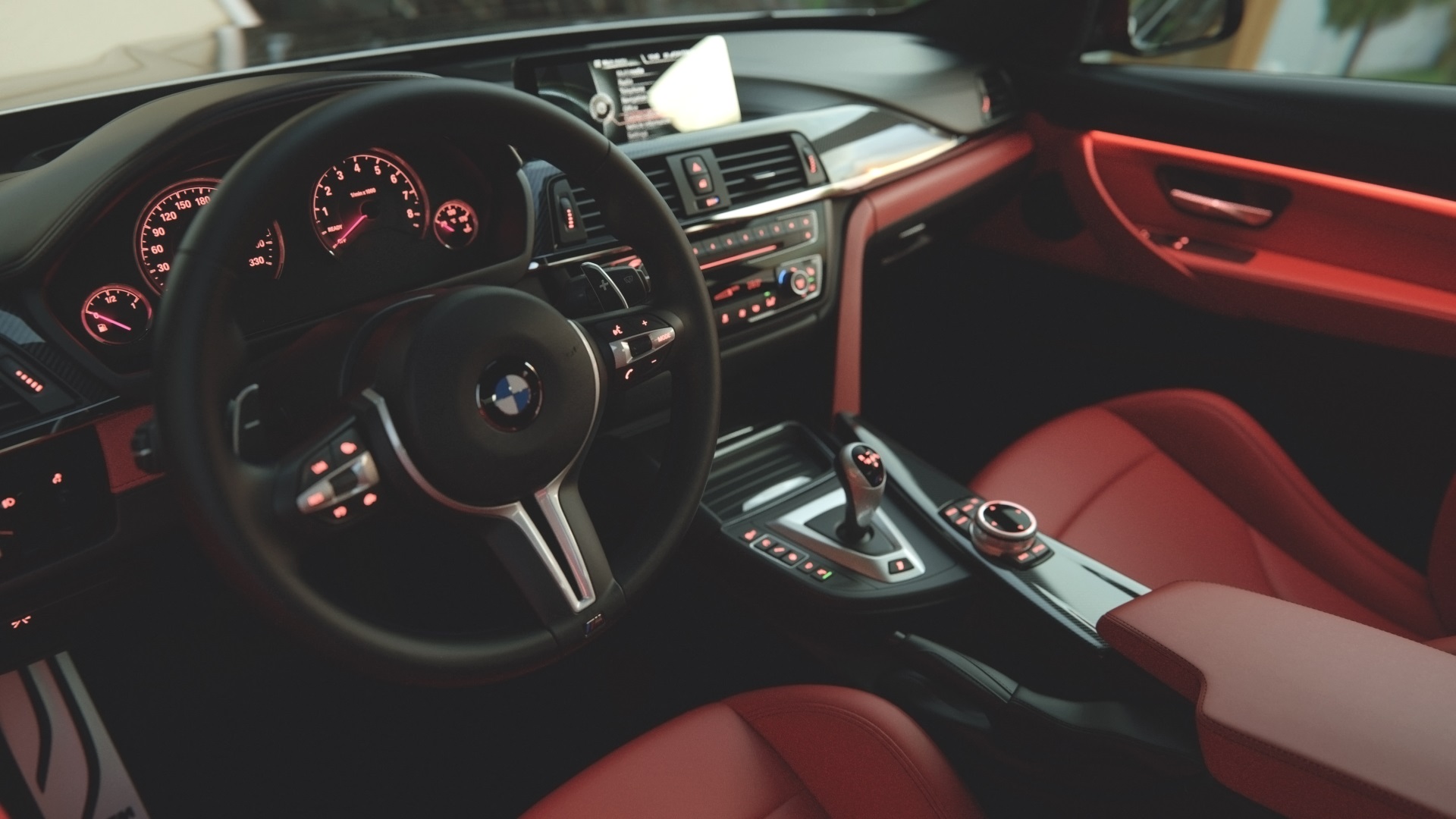} \hfill
	\includegraphics[width=0.4915\linewidth]{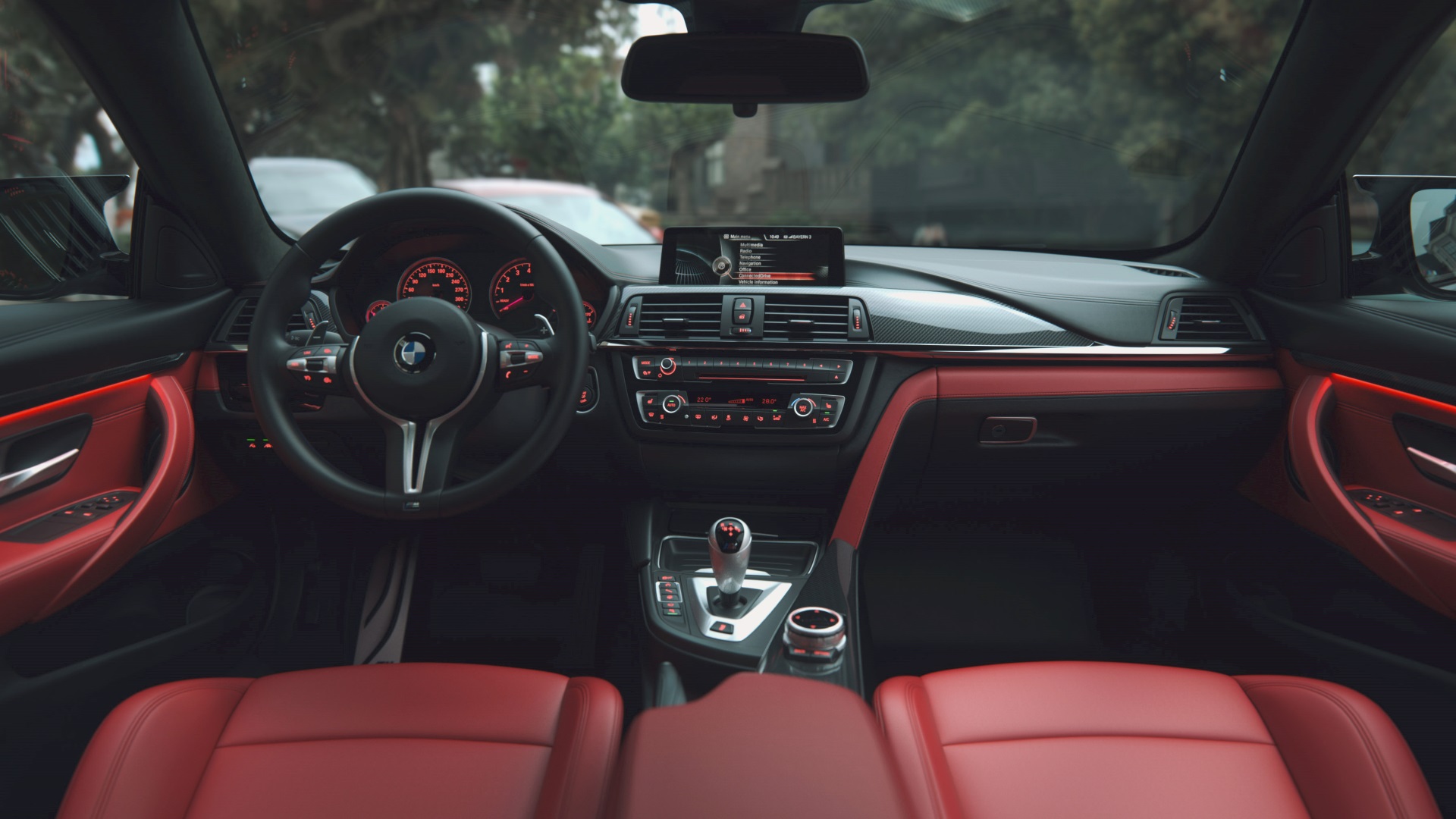}
	\caption{Car interior featuring subtle lighting effects from many textured geometric light sources on the dashboard and knobs. Images courtesy of Jeff Patton.\label{Fig:BMWinterior}}
\end{figure}

\subsubsection{Environment Light}

Importance sampling of the environment is accomplished by a hierarchical representation, the finest level of which is essentially an equirectangular projection of the directional intensity onto a rectangular image.
Coarser levels are downsampled versions at lower resolution, computed during a preprocess.
Values must be scaled by their respective areas on the unit sphere to account for distortion near the poles.
This image pyramid can then be traversed probabilistically, such that at each level one region of the next level is chosen according to its average intensity.
Once we reach a single pixel on the finest level, we can obtain the associated direction.

While this scheme already efficiently importance-samples the environment over the sphere of all directions, it does not take surface orientation and normal into account.
When used without modification, this leads to very inefficient sampling for shading points that face away from a high-intensity part of the environment, such as the sun.
Using a large number of individual hierarchies, each with its values scaled according to one of a set of discretized surface normal orientation would solve the issue, but is generally prohibitive in terms of preprocessing time, memory requirements and data coherency during the simulation itself.
Instead, we just provide the first few levels of the hierarchy in this fashion, approximately weighted according to surface normal orientation.
This is a unique advantage of the proposed multi-scale 2D representation over a simpler 1D method, such as an alias map, over the set of all environment map pixels.
Although the weighting for the top levels is done quite conservatively to ensure coverage of the whole relevant domain for the actual normal as opposed to the discretized one and only very shallow top-level hierarchies are built, this scheme typically can still provide very good irradiance distribution importance sampling at negligible memory overhead.

To deal with custom, user-programmed, MDL environment functions, Iray utilizes a fully automatic %
baking process to an environment map (see Fig.~\ref{Fig:Procedural}).
As an exception, a native sun and sky model is implemented in the core, which has explicit built-in evaluation and importance sampling support.

To aid the illusion of having a ``real'' environment surrounding the computer graphics (CG) scene, Iray offers
finite sized environment domes with box or sphere geometry, realized by a
projection of the infinite environment onto those shapes.
This is especially important when interacting with a scene interactively to get a reasonable sense of depth and size, which is missing completely when using the common infinite environment
setup.
We further offer an optional projection to realize a flat
ground plane with configurable distortion for the lower hemisphere.
In order not to introduce artifacts in conjunction with MIS-weighted rays hitting the environment, it is important to account for the dome shape in light importance sampling, too.
We require the projection function to be invertible within the dome,
such that importance sampling can still be performed on the infinite environment and
then be transformed to the finite dome. Compared to potentially supporting artificial geometry instead,
this is also much faster during runtime and shortens preprocessing.
Note that to obtain a correct result, the measure change from the
transformation to the projected domain needs to be taken into account
for the probability density and contribution computation.
While this yields some small amount of additional variance, it is
significantly simpler than performing importance sampling on the
finite domes directly.

\begin{figure}
	\centering
	\includegraphics[width=\linewidth]{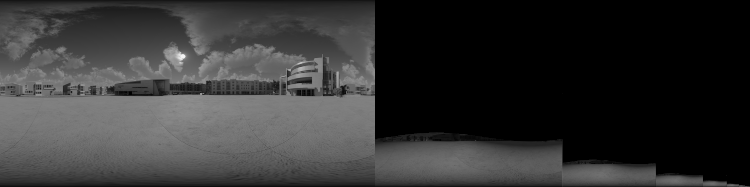} \\\vspace{2mm}
	\includegraphics[width=\linewidth]{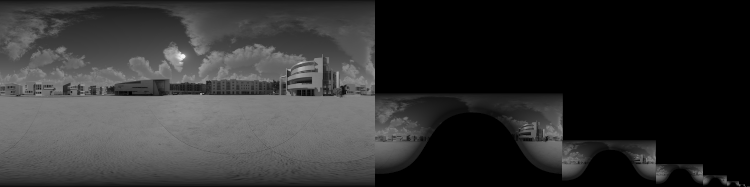}
	\caption{Two different top level hierarchies pointing into the same base level of an environment light. Values are re-mapped to a visible range to show the differing probabilities on the top levels. \label{Fig:EnvMap}}
\end{figure}

\subsubsection{Photon Aiming}

Iray uses light tracing in order to accelerate the rendering of caustic effects.
Light tracing generates light transport paths starting on the light sources, simulating the 
trajectory of photons and explicitly connecting the paths to the camera at every bounce. 
As such, a high photon density in screen space translates to fast convergence. 

However, traditional photon sampling methods do not take screen space photon density into account when generating photons.
Photons originating from an infinite environment are aimed uniformly towards a bounding volume of the entire scene. 
Similarly, photons originating from finite light sources are emitted proportional to the emission distribution of the light 
source \cite{Walter:1997:GIU:256157.256158}. Because screen space density is ignored, caustics close to the viewer and from far away 
light sources tend to be undersampled, while caustics in the distance or from light sources near to the caustic effect may be 
oversampled. 

This method is highly sensitive to scene setup and violates our push button design goal.
As an example, a very common scene setup in the design space is a closeup of an object (such as jewelry or cars) on a turntable, lit by an environment or several faraway light sources.
Simple scene modifications such as increasing the turntable size or adding any object far from the viewer will increase the total scene size and 
reduce the photon density on the object of interest, although the visual portion of the scene is not affected. 
Similarly, lighting the turntable with a faraway light source will result in very low photon density on the object of interest.
Many users will experience this as counter-intuitive.

Different methods have been proposed to improve the distribution of emitted photons. 
Jensen proposed the use of projection maps \cite{Jensen95-PMBMC}; a discretized map of emission directions on each light source. 
All potential caustic casting objects are projected back onto the map and then
all covered cells in the map are marked for photon aiming. During photon generation photons are aimed only at the covered cells, thus focusing photons on caustic casting objects only. 
This method however makes no distinction between the importance of different caustic casting objects. 
In practical scenes most materials have one or more highly specular component such as a glossy coating layer, therefore the projection maps will cover most of the scene and thus make little difference.

Peter and Pietrek \cite{Peter:1998:IDC} proposed to construct a piece-wise linear importance function in a preprocess, first constructing an importance map using regular path tracing and then estimating the importance function on the light source surfaces using importance gathering and piece-wise linear discretization for the estimate. 
A downside of this method is that it relies on path tracing and photon gathering techniques to construct an importance distribution estimate. 
As photon aiming is most useful in hard scenes where path tracing or plain photon emission fail of capturing the caustics,
relying on these same techniques to construct an importance distribution estimate can be considered impractical. 
Either many paths are needed in the preprocessing phase or the estimates will be very noisy, resulting in unreliable photon aiming distributions. 

Several Metropolis based methods also have been proposed to improve the photon distribution \cite{Veach:1997:MLT, CKelemen, Hachisuka:2011:RAP:2019627.2019633}. 
These methods still rely on pure photon tracing to coarsely explore the path space and find caustics. 
When a caustic path is found, its neighborhood is further explored by making small mutations to the found path.
However, these methods still greatly benefit from a good initial photon emission distribution to reduce the correlation between paths, typical for all Metropolis based rendering algorithms.
 
In Iray we use an alternative method for improving the photon emission distribution by aiming photons non-uniformly at scene objects. 
To fit our push-button design goal the distribution over scene objects is constructed automatically without user assistance.
The algorithm constructs a static sampling distribution for a given scene and camera position, contrary to an online exploration based method such as Metropolis.
Our main motivation here is scalability, and online exploration methods are inherently sequential which severely limits scalability (see Sec.~\ref{Sec:Parallel}).
Based on the assumption that the photon density of caustics in screen space is strongly correlated with the screen space area of photons on caustic-casting objects,
which is typically true for reflective and refractive caustics in product design scenes, caustics tend to show up in the near vicinity of the reflecting/refracting objects that cast them.

Our approach works by automatically clustering objects and aiming photons at these clusters proportional to a weight which estimates the expected screen space photon density. 
When emitting a photon from a light source or the environment, a cluster is stochastically selected proportionally to its weight. 
The photon is then aimed uniformly at the bounding volume of the selected cluster. 
This method then robustly simulates caustics in open scenes with large ground planes and several nearby and far away objects of varying 
sizes, even when the caustics are caused by a combination of nearby and faraway light sources. 

Cluster sampling and probability density evaluation is sped up by storing the clusters in a bounding volume hierarchy, similar to 
the acceleration data structure for geometric light sources. 

The objects are clustered in a bottom up fashion by repeatedly merging nearby object clusters with similarly-sized object clusters, along 
the lines of \cite{walter08rt}. Initially, each object forms a single cluster. Clusters are considered similar if the smallest object within 
either cluster is not much larger than the complete other cluster. This method tends to find all obvious clusters without generating 
many overlapping clusters of similar size. Another advantage of the clustering criterion is that the final set of clusters is independent 
of the order in which clusters are merged, thus giving more stable and predictable results. 

A cluster is sampled by traversing the cluster hierarchy from the root to a leaf. 
At each inner node, one of the children is stochastically selected proportionally to an approximation of the screen space photon 
density from the children. The photon density of a node is estimated from the position of the camera, the light source, and the 
bounding volume of the node. In particular, the photon density is estimated as the highest screen space photon density of any 
point in the node's bounding volume. The optimistic estimate prevents singularities or oversampling of clusters that cannot be 
sampled efficiently anyway. Similarly, the evaluation of the aiming probability density for MIS is sped up by intersecting the emission ray 
against the bounding volume hierarchy to find all overlapping clusters and compute their sampling probability density.

\subsection{Material Evaluation} \label{Sec:MatEval}\label{Sec:MDL}

\begin{figure}[t]
	\centering
	\includegraphics[width=\linewidth]{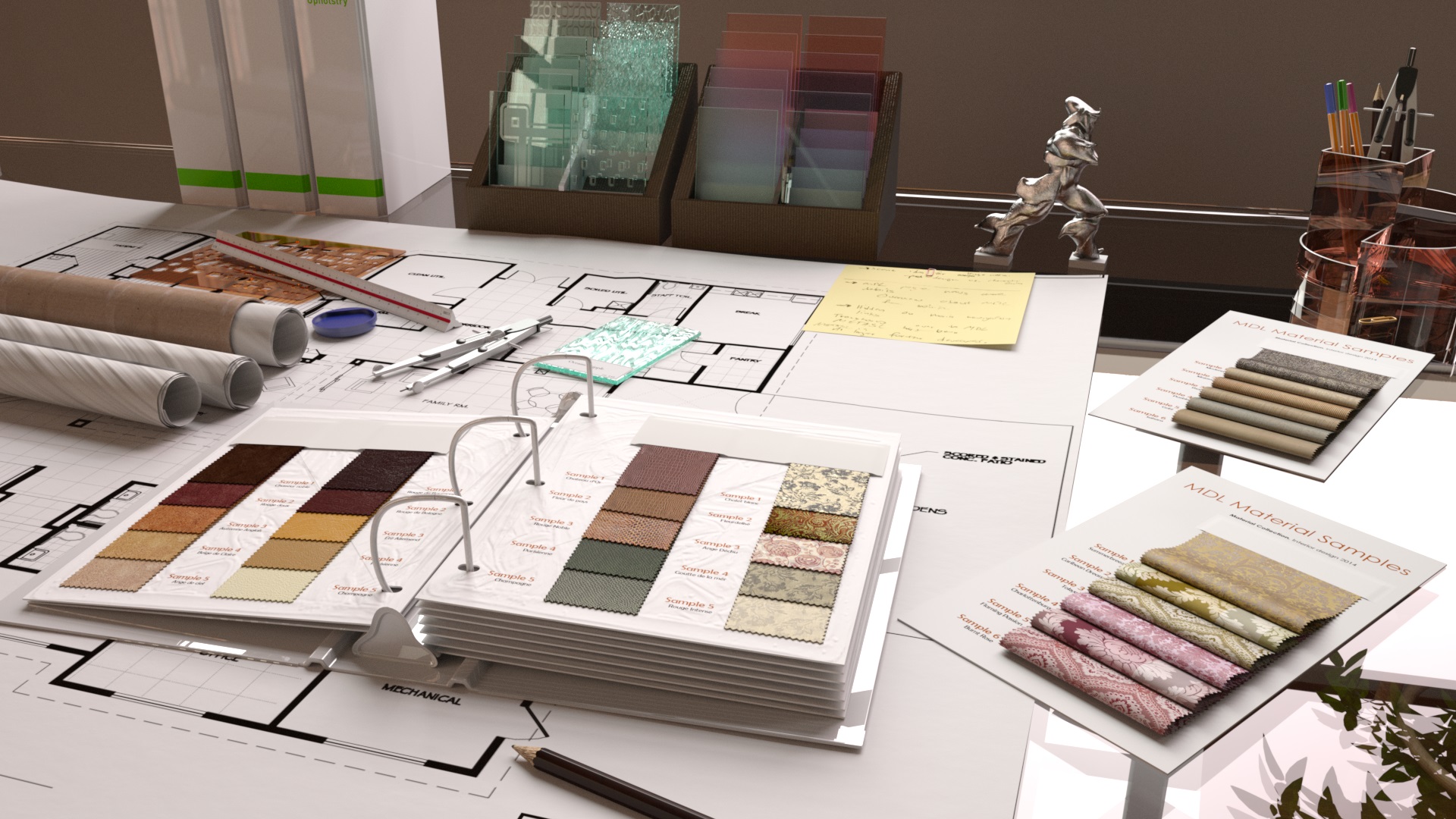}
	\caption{Virtual material catalog designed with MDL.\label{Fig:architect_studio}}
\end{figure}

A key aspect in the design of the Iray system is a material model that can be both efficiently evaluated and importance sampled.
To be of any use for advanced rendering, and in particular to compete with user-programmable shading, this material model needs to be highly flexible and must offer a great degree of artistic freedom.
MDL \cite{MDL} was designed alongside Iray's material model for exactly that purpose. Fig.~\ref{Fig:architect_studio} shows some examples of materials designed with MDL.

\begin{figure}[t]
	\centering
	\includegraphics[height=0.3\linewidth]{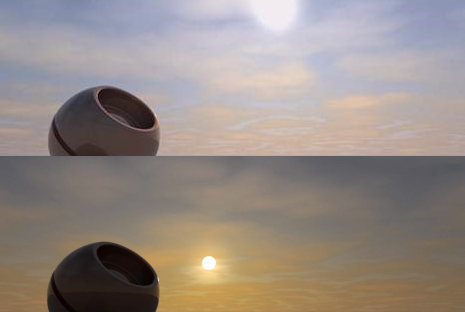} \hfill
	\includegraphics[height=0.3\linewidth]{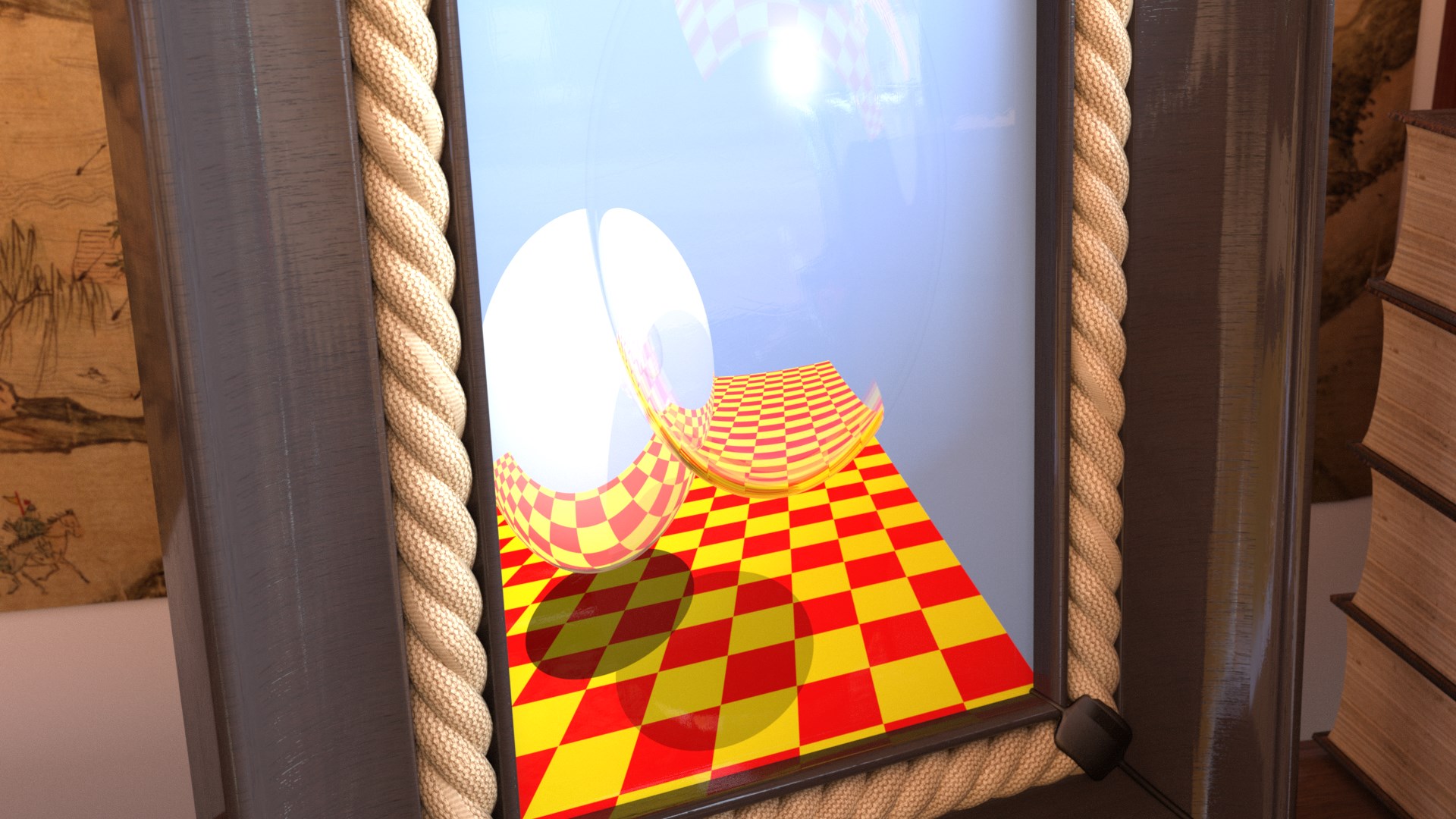}
	\caption{Left: User-programmed sky model with two different sets of parameters (Top, Bottom). 
		Right: A procedural texture (inside the picture frame) which implements a simple ray tracer \protect\cite{Whitted:1980:AII} completely in MDL \protect\cite{WhittedMDL}. \label{Fig:Procedural}}
\end{figure}

To give a brief overview, MDL is a domain specific language that is divided into two parts:
\begin{enumerate}
\item
  The declarative part describes the fundamental properties of the
  material. In particular, a hierarchically layered bidirectional scattering distribution functions (BSDF) model allows for flexible combinations of
  a rich set of common elemental BSDF models, including measured datasets. 
  This declarative part also includes emission, volume coefficients, and geometric properties, such as index of refraction
  and main shading normal.
\item
  The procedural programming language allows free programming of
  material inputs, such as BSDF color tint and per layer bumps.
  Functions defined in the language are side-effect free and operate on a small, read-only
  rendering state to allow for efficient execution on current hardware architectures.
  Note that while certain common basic functionality like bitmap lookups,
  texture blends, and procedural noise functions are implemented as optimized core code,
  in general any user provided MDL function is supported by
  the means of just-in-time compilation for NVIDIA GPUs and CPUs
  (see Fig.~\ref{Fig:Procedural}).
\end{enumerate}

Due to the large degree of user flexibility that MDL affords, material evaluation is the largest execution stage of Iray.
The core of this stage is comprised of BSDF evaluation and importance sampling, but several other important aspects contribute as well
and a large amount of data needs to be accessed at various points during the execution.

We can think of the material evaluation as separated into four parts:
\begin{enumerate}
\item State setup,
\item texture coordinate and tangent generation,
\item computation of material inputs including bitmaps, procedural textures, and compiled MDL code, and
\item evaluation and sampling of the layered BSDF.
\end{enumerate}

The first step includes setting up the local intersection information like front- and backside point of intersection (to handle 
both reflective and transmissive events properly and without geometry self-intersection issues) and the geometric and the interpolated shading normals.

This is followed by the processing of texture coordinates and corresponding tangent spaces, as needed by the material.
In general, 3D texture coordinates may be generated procedurally either from scratch or by transforming existing coordinates.
Since the various material inputs can require different sets of texture coordinates and tangent spaces, this data is computed upfront
and stored in the state to be picked up in the third stage, which avoids multiple computations of the same data.

The material input evaluation in the third step is performed in batches to improve efficiency. 
The texturing node graph of inputs for each material is sorted according to type, such that code and data divergence during evaluation are reduced.
This is especially beneficial in an SIMD or SIMT execution model, where multiple threads are more likely to operate on the same code and data at a time. 

Since inputs can depend on each other, special care has to be taken to resolve dependencies in this sorting step, 
such that intermediate results are already available.
The sorted batches are then processed and the results are picked up by the hierarchically layered BSDF code.

Finally, the BSDF evaluation and sampling step is implemented in a generic way such that any hierarchical layering defined in MDL can be processed
by the same code. In contrast to specialized code per material, as created per just-in-time compilation, this again reduces divergence.

\subsection{Volumes} \label{Sec:Volumes}

Besides light interacting with the material surface, Iray supports homogeneous volumes. %
Contrary to many rendering systems, Iray does not feature dedicated approximations
for simulating sub-surface scattering effects. This option was deliberately chosen to
have significantly simpler code, higher robustness for arbitrary geometry (convex and highly
detailed regions), and a general solution for nested volumes at the same time, in keeping with the idea of preserving generalization.

To support nested participating media it is important to offer a simple
scheme to model such volumes, without the need for additional flags or
priorities. For that reason, Iray implements an extended stack to
store a reference to the current material including its volumetric properties.
To handle the transition from one volume boundary to the other in a robust and precise manner
it is sufficient to model the volumes with a slight overlap which is automatically handled by the
stack traversal code. This avoids the common problem of non-matching tessellation levels for
neighboring volume boundaries or general ray tracing precision issues \cite{Woo:1996:IRN}
which can result in small ``air''-gaps or missed volume transitions that falsify all refraction and absorption computations.
Our stack traversal code filters the volume transitions based on the information on the stack
to avoid that the overlapping volumes trigger multiple media changes instead of just one.
Note that the simple case of a fully enclosed volume does not require special treatment.

\begin{figure}[t]
  \centering
  \includegraphics[width=\linewidth]{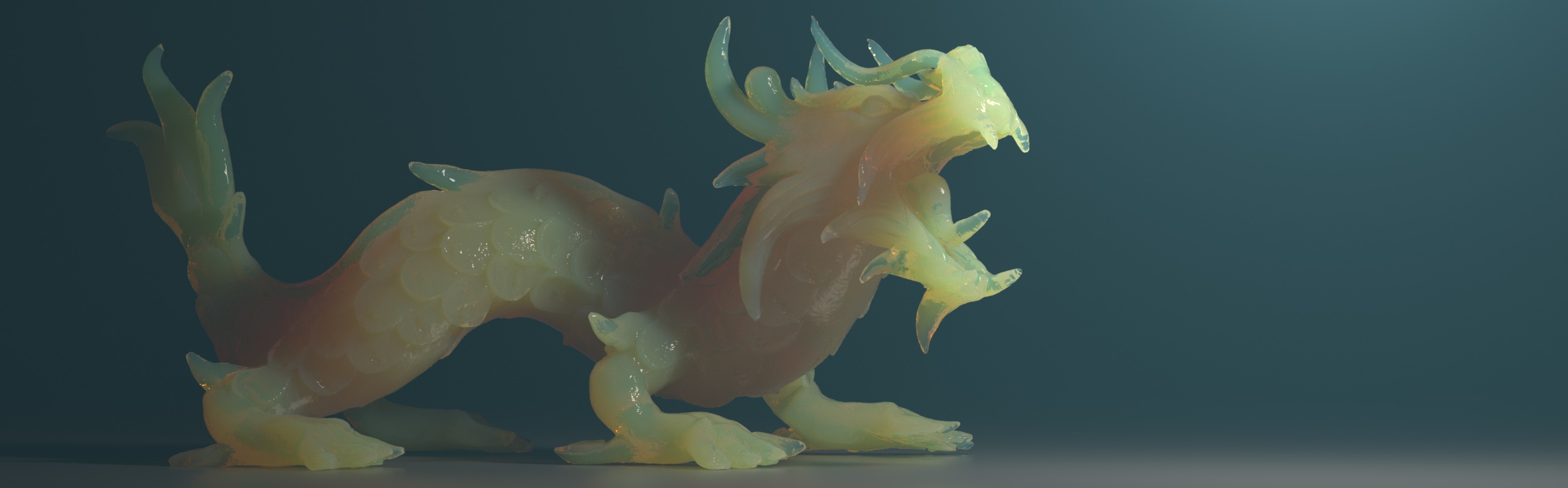}
  \caption{Scattering in participating media, showing two nested volume objects with different densities. The light fog of the outer volume (containing both camera and light source) appears blue due to scattering coefficients chosen to approximate Rayleigh scattering. Dragon dataset courtesy of XYZ~RGB and the Stanford 3D scanning repository.\label{Fig:Volume}}
\end{figure}

As the camera itself can be inside of a nested volume (see Fig.~\ref{Fig:Volume}), it is necessary to initialize the stack
accordingly. A preprocessing step connects the camera with the bounding box of the scene
to fill in the volumetric properties of all surrounding media. Since the correctness of this preprocessing is essential for
the following runtime computations, special care has to be taken for the ray tracing computations to ensure
that no volume interactions are missed. 
Iray achieves this by using watertight hierarchy traversal and triangle intersection algorithms everywhere.

\subsection{Spectral Rendering}

Iray supports both spectral and colorimetric rendering within a
single rendering core, implemented in a way that neither
negatively impacts performance nor sacrifices generality by introducing special implementations.
Internally, the system is largely color-agnostic and merely operates
on triplets of values and only very few specialized code paths exist that
actually distinguish between spectral and colorimetric input.
Spectral rendering encompasses the complete simulation, including texture input, lights, surface material properties, and volume coefficients.
If spectral rendering is enabled, up to three wavelengths are
simulated simultaneously, each one importance sampled (typically
according to the CIE XYZ color matching functions) using equidistant
offsets along the spectral dimension (similar to \cite{WNDWH14HWSS})
and weighted using spectral MIS \cite{spectralMIS}.
By default, the result of spectral rendering is converted to color,
but it is also possible to accumulate unweighted averages for up to
three wavelength bands.

\begin{figure}[t]
	\centering
	\includegraphics[width=\linewidth]{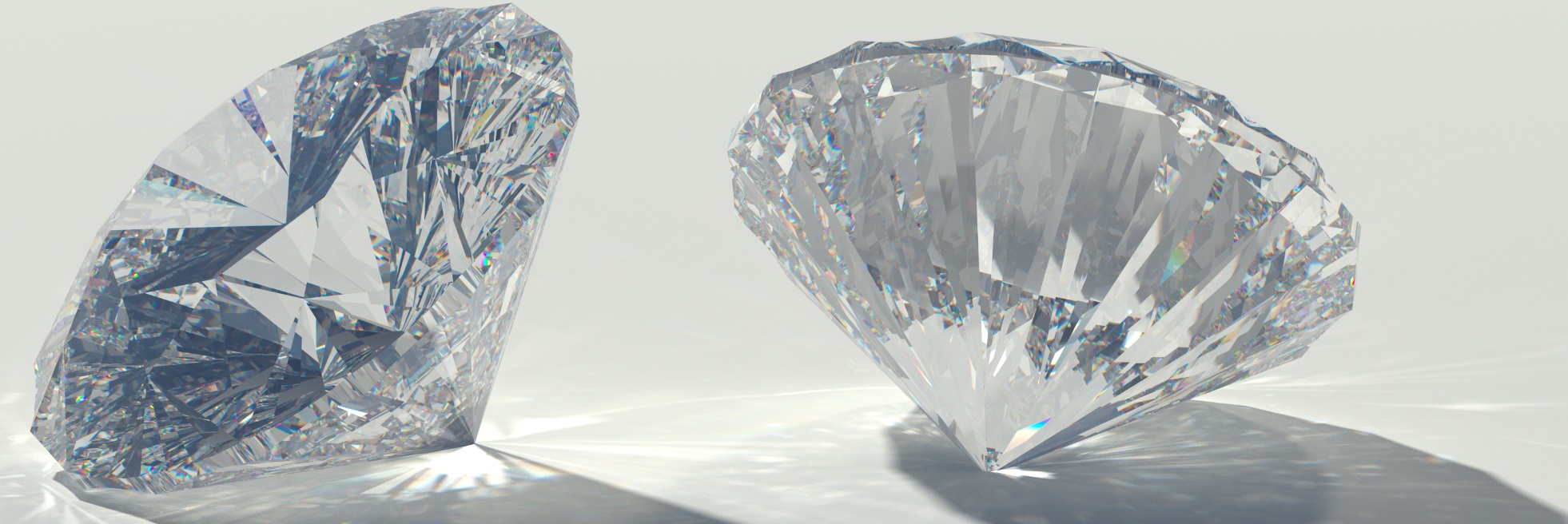}
	\caption{Dispersion effects without full spectral rendering. The material is defined only by applying the measured index of refraction and Abbe number of diamond, both taken directly from the literature. Diamond scene by Paul Arden.\label{Fig:Dispersion}}
\end{figure}

Spectral rendering in Iray is intended to be a transparent runtime
switch with as few implications on the user as possible: both spectral
and colorimetric rendering support spectral and color input data, with
automatic conversions in place.
For the conversion process, the goal is to match the ideal result closely, e.g.\ if color data
is used for spectral rendering then the final result should ideally not
differ from the corresponding colorimetric rendering.
While in general, of course, this is not possible, a
good match for direct lighting could be achieved by knowing the dominant spectral illuminant
and incorporating this information into the reflectance
conversion.
As in general this is unknown, we base the conversion on the assumption that the corresponding
spectral illuminant is the spectral white point of the color space
(e.g.\ D65 for Rec.709 and Rec.2020, D60 for ACES).

\begin{description}
\item[For spectral rendering,] color input (including bitmaps) is converted to
  spectral at runtime.
  Conversion is based on chromaticity coordinates along the lines of
  \cite{spectralMeng}, however, using a different construction for the table data:
  For emission we require that $(1,1,1)$ yields the spectral white
  point of the source color space and reflectance 
  spectra are optimized to be smooth relative to that spectrum.
  Optionally, a variant of \cite{spectralSmits} can also be used for
  reflectance, which, at the cost of a less smooth spectrum, allows one to
  reconstruct all fully saturated reflectance colors within the Rec.709
  gamut without energy loss.
  
\item[For colorimetric rendering,] spectral input (including spectral bitmaps) is
  converted to color before rendering, where for reflectance the conversion
  is similar to \cite{spectralWard}.
\end{description}

Spectral rendering capabilities are sometimes motivated by the desire to render spectral dispersion, i.e.\ rainbow patterns caused by a material's index of refraction varying with the wavelength.
However, we believe that spectral rendering yields little advantage over a color-only pipeline whenever this is the only motivation.
In fact, Iray had supported plausible spectral dispersion long before featuring true spectral rendering:
Locking the ray to a sampled wavelength at the first dispersing material hit and tinting the color accordingly will give a result that is often indistinguishable from the correct solution at negligible additional cost (see Fig.~\ref{Fig:Dispersion}). %
We see the real benefit of spectral rendering in the fact that it removes implicit approximations that come with multiplying tri-stimulus colors (light and material), thus resolving artifacts and issues with e.g.\ metamers.
Therefore, assuming the availability of spectral input, it enables the user to create a truly predictive result, even for secondary bounces and in mixed lighting situations. %

\subsection{Motion Blur} \label{Sec:MotionBlur}

\begin{figure}
	\centering
	\includegraphics[width=\linewidth]{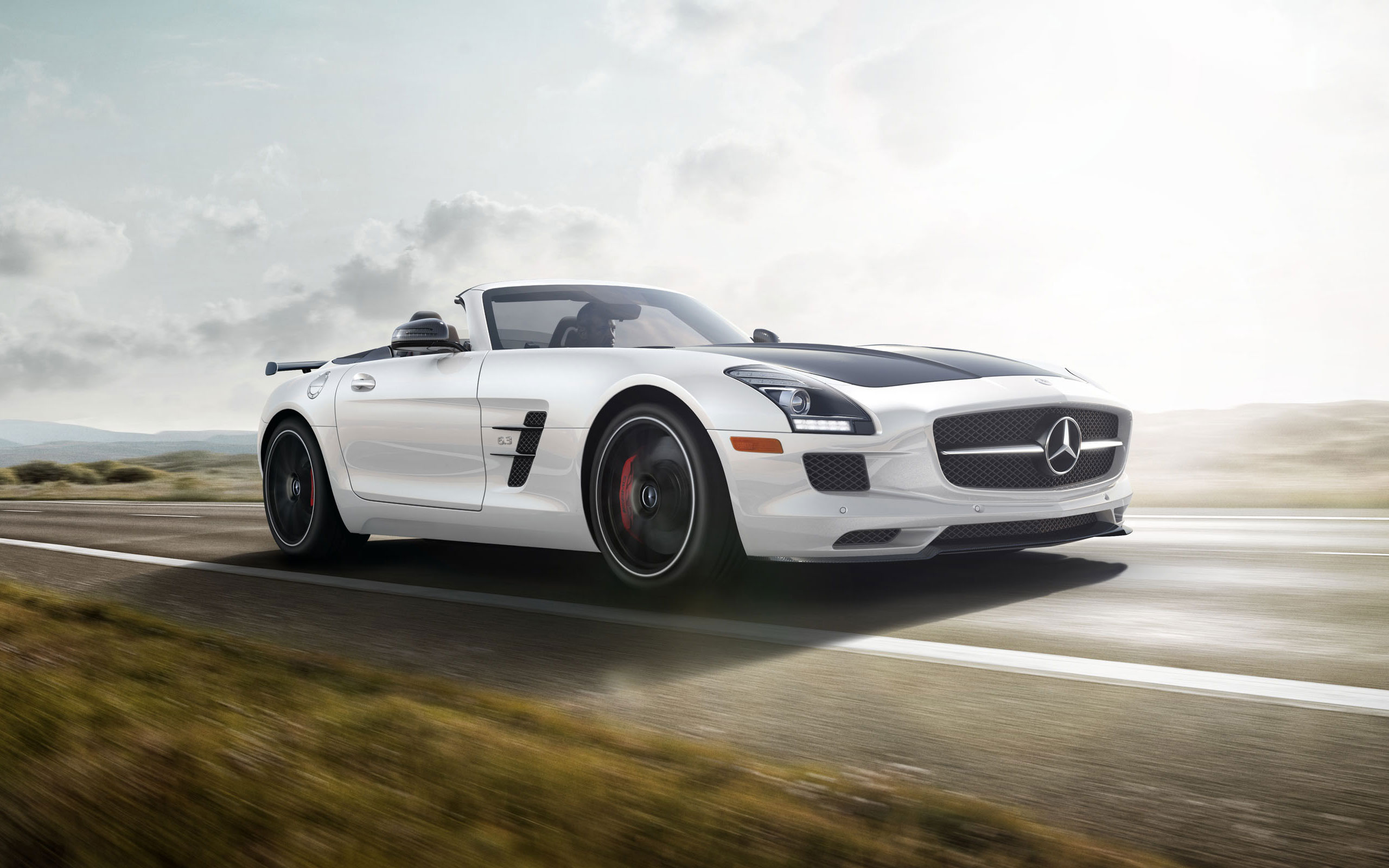}
	\caption{Motion Blur, using the sampled transformation. Image courtesy of MBUSA (rendering and setup by Jeff Patton). \label{Fig:Car1}}
\end{figure}

Contrary to common approaches that incorporate the concept of motion
at the core of their implementation, Iray uses neither motion vectors nor interpolation, but the real, sampled transformation.
The scene data for the chosen exposure time of the virtual camera or measurement probes is sampled for each iteration, which then works on a single point in time of the simulation.
This improves the quality of all the underlying acceleration data structures and avoids various
problems and artifacts caused by the interpolation of time-varying geometry, attributes,
and material inputs (see Fig.~\ref{Fig:Car1}). It also simplifies a lot of the core code that otherwise
would have to deal with handling the interpolation of transformations and motion vectors.
This approach is made practical by the rapid GPU
construction of the acceleration data structure for ray tracing \cite{Karras:2013:FPC:2492045.2492055} (also see Sec.~\ref{Sec:Parallel}).

\section{Scalable Parallelization} \label{Sec:Parallel}

Despite all importance sampling optimizations, light transport simulation often
still requires a vast number of path samples to render noise-free
images and, therefore, is very amenable to massive parallelization.
Besides using GPUs,  where a wavefront approach (see Sec.~\ref{Sec:StateMachine}) is used to extract parallelism 
from a large set of concurrently generated light transport path samples,
Iray is designed to scale across the range from a single GPU
or CPU to a heterogeneous cluster of hosts, each with multiple GPUs (see Fig.~\ref{Fig:schedflowchart}).
In order to be able to combine intermediate results, the simulation must produce numerically close results on all supported platforms.
Accordingly, the vast majority of Iray's kernel code is written in an abstracted hardware-agnostic way and shared as-is for both CPU and GPU.

Geared towards two different use cases, 
Iray implements two strategies of parallelization:
During phases of user interaction (see Sec.~\ref{Sec:Interactive}) the latency of
progressive updates is minimized, while in batch mode 
a single converged image is computed as fast as possible (see Sec.~\ref{Sec:Batch}).
Switching from interactive to batch mode while rendering is seamless without
losing any rendering progress.

\subsection{Parallelization on a Single Device} \label{Sec:StateMachine}

Each block of the state diagram in Fig.~\ref{Fig:sampleflowchart} can be implemented as a function that
transforms the sample from one state to the next.

Such functions may allude to implementing the state
machine as a so-called megakernel, i.e.\ by just constructing all
samples in SIMD fashion in a single kernel. While this approach certainly can be
successful with simplistic material models \cite{BikkerPhD},
it has been shown that on GPUs queueing samples by state (see Fig.~\ref{Fig:schedflowchart})
and executing individual state kernels on batches of
queued samples as a wavefront can be much more efficient \cite{Wavefront}.
But in order to avoid excessive persistent rendering state or loading (and initializing) data multiple times, the separation into
different states is always based on a trade-off, which is why for example, the material evaluation stage is not further subdivided.

In addition, the Iray rendering system uses a hybrid approach, switching between wavefront and megakernel
implementations based on the queue size (see Sec.~\ref{sec:regeneration}).

\begin{figure}
  \centering
  \includegraphics[width=\linewidth]{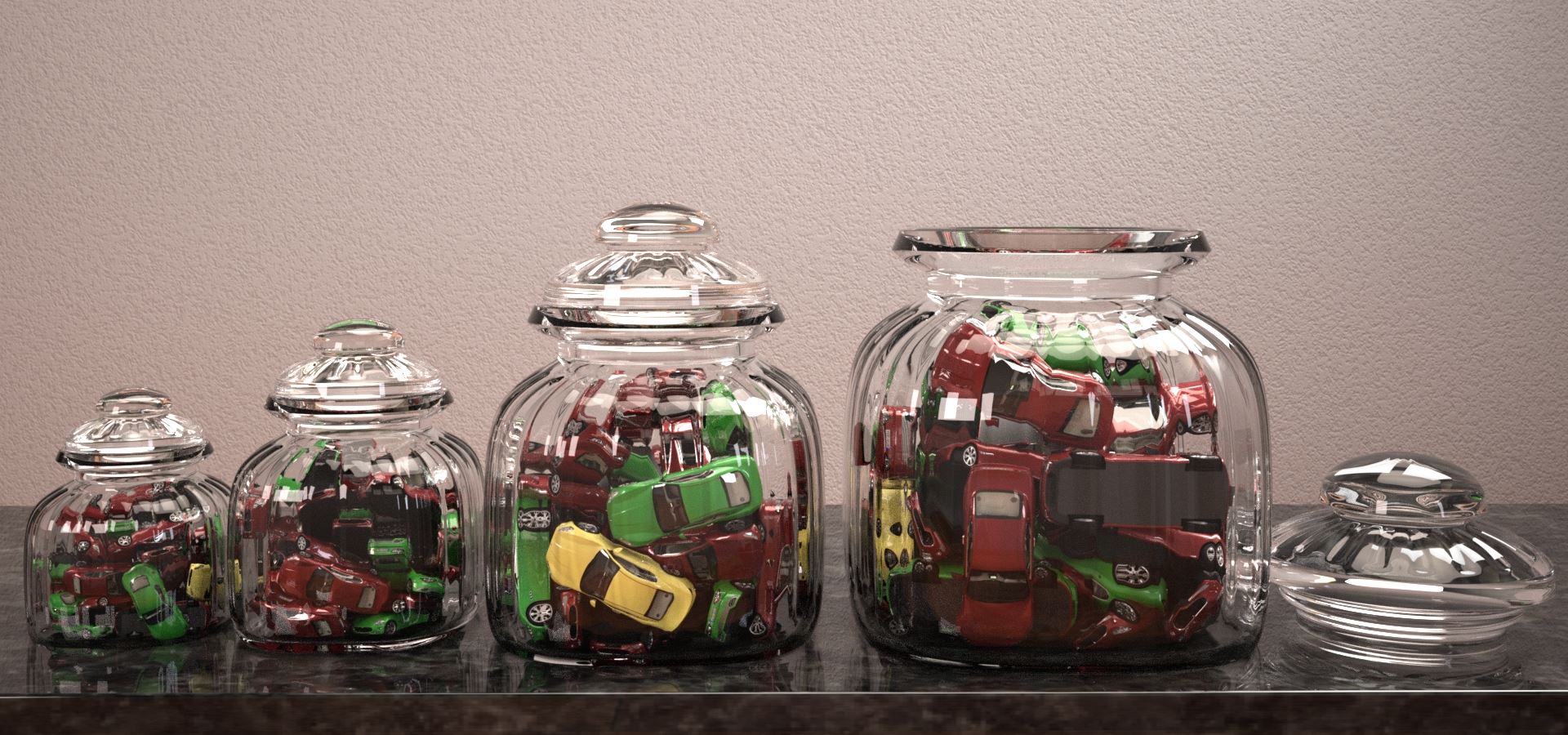}
  \caption{141 instanced cars. Each car consists of roughly 1.4 million triangles. \label{Fig:Instancing}}
\end{figure}

For the ray casting state kernels our system employs the low level ray tracing API OptiX Prime which comes
with the NVIDIA OptiX ray tracing engine \cite{OptiXPrime:2014}.
Fig.~\ref{Fig:Instancing} shows one use of instancing, enabled by OptiX Prime.

\subsubsection{Sample State}

During construction of a sample the complete state of the light transport path is stored in a sample state data structure.
This data structure has an uncompressed size of about 1~kB, reduced to about 0.65~kB when compressed. While this considerable size may be surprising, it is important to note that volume stack (see Sec.~\ref{Sec:Volumes}), light path expression state (see Sec.~\ref{Sec:LPE}), next event estimation data (maintaining a separate state to connect to the light sources), multiple importance sampling (MIS) state, and matte path state (see Sec.~\ref{Sec:matte}) need to be stored for simulation. 
The size has already been reduced by combining data with exclusive lifetimes in unions.
Although most data is only used by a subset of all state kernels, their lifetimes usually overlap, spanning multiple scattering events.

For the state machine execution, a fixed sized array in the order of a million sample states is allocated in GPU global memory.
Storing the sample states as an array of structures (AoS) results in divergent and non-coalesced memory accesses whenever threads running in SIMT access the same field in adjacent sample states, severely reducing performance. 
Laying out the data as a structure of arrays (SoA) roughly doubles Iray's overall performance.

For compressing unit vectors we used the octahedron mapping flattened to a 2D layout \cite{OctEnvMap} with a 16-bit-quantization for each component (so achieving an overall 3:1 compression ratio). 
For unit floats a uniform quantization to 16 bits is used (2:1 compression ratio). 
For all floats that suit a limited range and limited precision, the half-precision format is used (again achieving a 2:1 compression ratio).
Using these compression schemes does not result in noticeable differences in the converged images.

\subsubsection{Sample Queue} 
\begin{figure}
  \centering
  \includegraphics[width=\linewidth]{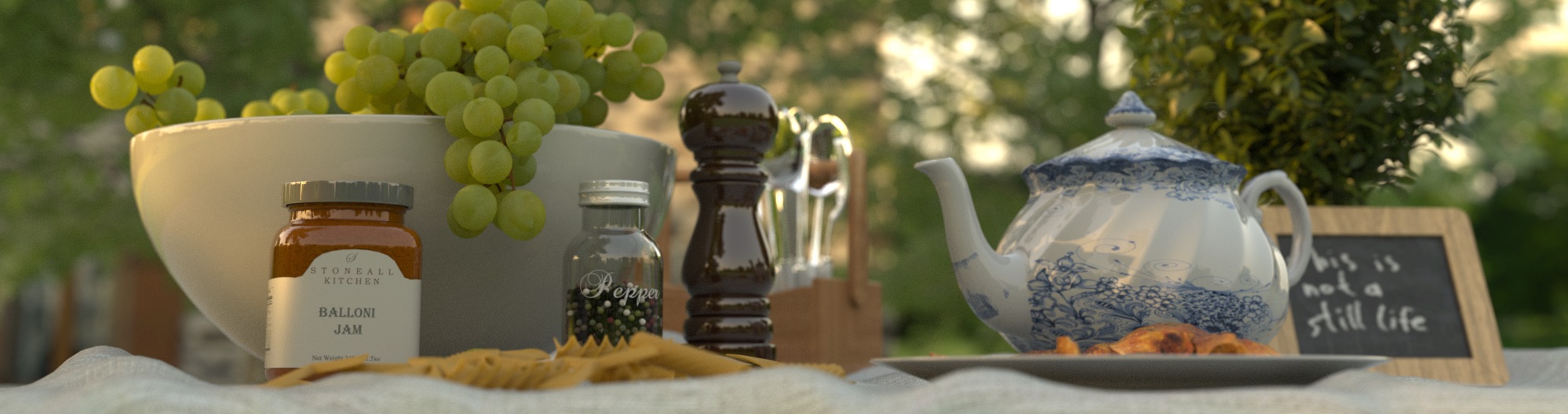}
  \caption{The Still Life scene uses a large variety of different material types to test the behavior of the simulation for incoherent workloads. \label{Fig:StillLife}}
\end{figure}

\begin{figure}[H]
  \centering
  \includegraphics[width=.7\linewidth]{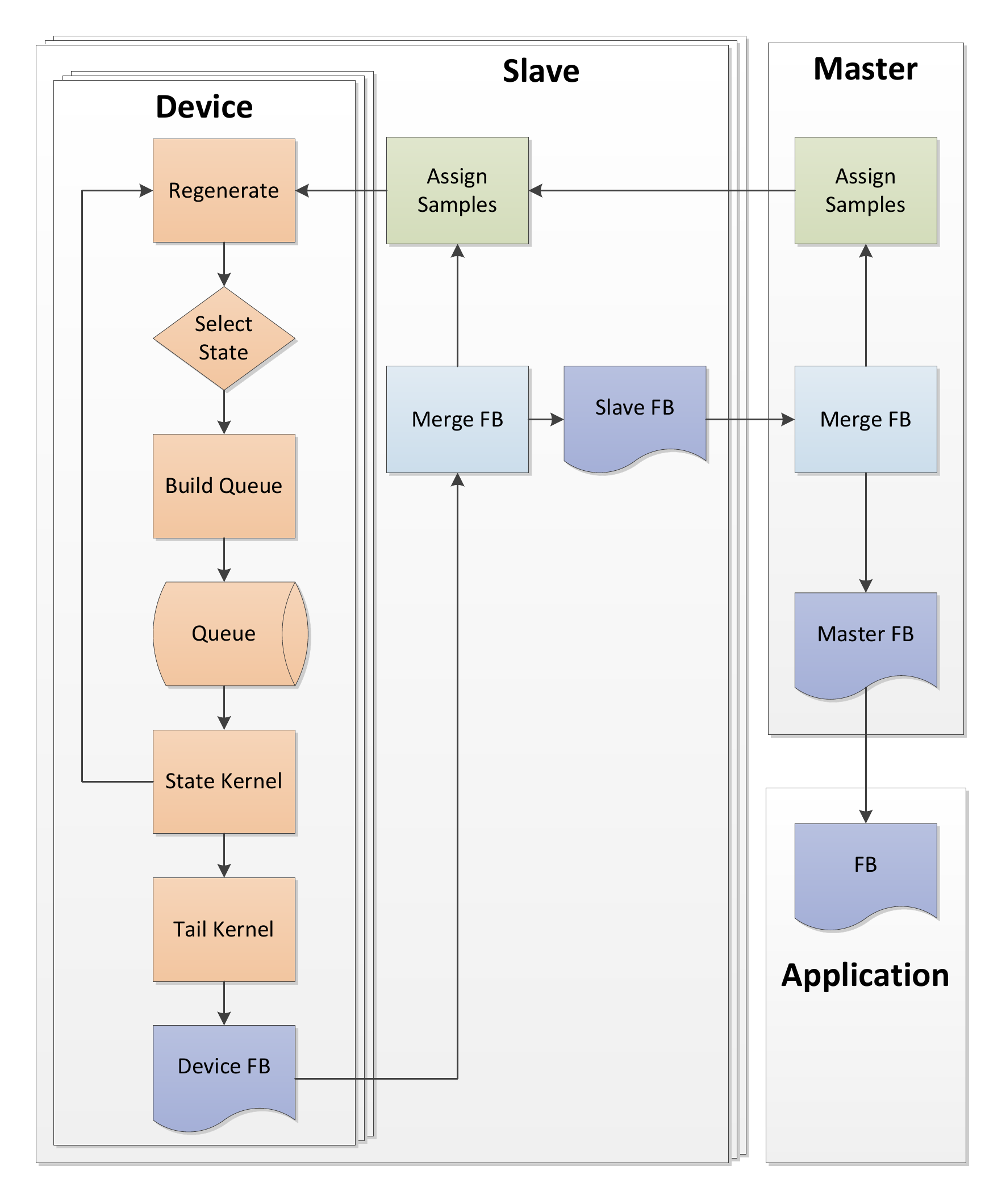}
  \caption{Distributing samples for parallelization
  and how the intermediate rendering results are merged into the framebuffer (FB).
  A master process controls multiple slaves that can host multiple
  compute devices, e.g.\ GPUs.\label{Fig:schedflowchart}}
\end{figure}

The state kernel to be executed next is selected in a greedy fashion, picking the most common state among all active samples (see Fig.~\ref{Fig:schedflowchart}).
Atomics are used to keep track of the number of samples in each state.
After selecting the next kernel for execution, a queue of sample state indices associated with the selected kernel is constructed by filtering the sample states and generating a compact array.
Because each kernel only touches a small subset of the state, the queue only contains sample indices while state data resides at a fixed location in memory.
Allocating only a single queue in a just-in-time fashion instead of maintaining separate state kernel queues reduces memory overhead.
Furthermore, regenerating a compact queue from scratch for every state kernel improves state access performance because the sample indices in the queues do not lose coherence over time. 
These benefits come at the cost of the additional compaction kernel. 
When the runtime of the compaction kernel exceeds the runtime of the state kernel, we generate a single tail queue once, which contains the indices of all remaining active sample states, regardless of their state.

\subsubsection{Creation and Termination of Sample States} \label{sec:regeneration}

As light transport samples may be terminated after a different number of scattering events and at different states, the
number of active sample states decreases over time. 
Note also that the greedy state selection policy may block some samples reaching a rare state until all other samples have completed.
This creates a tail where sometimes even as much as 30\% of the render time of a single frame is spent on the last 5\% of
remaining sample states. This is due to the overhead of the kernel
calls for a handful of sample states and the resulting low GPU utilization
of the kernels. To address this issue, we use two methods.

First, we simply
switch to a megakernel approach once the number of remaining active sample states
becomes too low. This megakernel implements the full state
machine and completes all remaining samples in one execution, leaving the scheduling solely to the hardware.

The second method regenerates terminated sample states \cite{Novak11GPUCG}:
In order to amortize the cost of additional kernel calls, a just-in-time queue
containing the indices of all terminated sample states is created
once more than half of all the sample states in memory have been terminated.
Then, a regeneration kernel is executed to initialize these sample states with new samples.

Note that regeneration only helps if a sufficient number of
sample states can be regenerated, which may not be the case
close to the end of a rendering task. Tab.~\ref{Tab:Efficiency} shows 
the relative performance of the wavefront and megakernel approaches and how 
regeneration improves the rendering efficiency of the wavefront scheme 
as the sample size increases. Although the wavefront approach consistently wins out on
the megakernel in the long run, the speedup and break-even point
vary wildly from scene to scene.

While both methods are not exclusive,
aggressive regeneration reduces the benefits of the
megakernel. As the megakernel is aimed at interactive rendering (see Sec.~\ref{Sec:Interactive}), where the
sample size per progressive frame is often too small and regeneration becomes ineffective,
sample regeneration is most beneficial for batch rendering (see Sec.~\ref{Sec:Batch}).

Due to the wavefront implementation, we were able to improve the performance especially on modern NVIDIA GPUs. While the 
benefits are moderate on the 8-year-old Fermi-based generation (measurements across our internal test suite show performance improvements in the 
1--1.5x range), the more recent Kepler architecture (with a range of about 1.5--2x), and especially Maxwell- and Pascal-based GPUs (featuring a 
range of mostly 2--3x improvement over the pure megakernel model) benefit the most.

\subsection{Balancing Load across Devices and Hosts}

Load balancing is a major concern when generating images on inhomogeneous
sets of compute devices (CPUs and GPUs in a machine, and between machines in a
network) in parallel. 
We are primarily interested in static load balancing schemes where all available
work is split up a priori and no (or very little) rebalancing is performed.
This avoids excessive communication overhead especially in cases where many
machines in a network participate in rendering images with few samples per
pixel, which is common in interactive applications. 
In addition, we want to support elastic scaling in a network by dynamically adapting to machines entering or leaving the cluster.

\subsubsection{Interactive Rendering}\label{Sec:Interactive}

In order to achieve optimal efficiency, the workload should be distributed 
such that each worker's load is proportional to its relative rendering performance. 
This is commonly done by splitting up the image in tiles or blocks of pixels and 
distributing these over the workers. Unfortunately, the cost of rendering a 
pixel can vary drastically between different regions in the image, so na\"{\i}vely 
splitting up the image can easily lead to an unbalanced workload distribution. 
Assigning scanlines to workers in round-robin fashion, on the other hand, 
generates more uniform workloads, but assumes that all workers have the
same performance.

We use a different workload distribution scheme to overcome these issues. 
Our method employs a simple deterministic low discrepancy
sequence to quasi-randomly permute a set of small stripes of pixels.
We choose a stripe size that maximizes the total number of stripes, while 
enforcing a size of at least 128 pixels (for cache efficiency).
The stripes are permuted using the van der Corput radical inverse in base 2 on the stripe set indices
which are then distributed between workers, proportional to each worker's
relative performance.

Consequently, the pixels that are assigned to each worker are well distributed
over the image.
For production scenes, the stripes are unlikely to correlate strongly with image
features, so that we can expect the computation cost between different sets of
stripes to roughly correlate with the number of pixels in the set.

The relative performance of each device is estimated at startup based on 
hardware specifications and refined during rendering based
on actual rendering performance. The framebuffer partition is continuously
rebalanced during rendering. If a device fails, the framebuffer is redistributed 
for the remaining workers and the failed samples are re-rendered.

Each worker allocates a partial framebuffer to hold its assigned stripes and
renders all pixels in the stripe assignment.
While rendering, the pixels are mapped to screen space using the permutation of
the stripe indices, which is its own inverse.
When rendering is complete, the stripes from partial framebuffers are scattered into
the final full framebuffer, again using the permutation. 

The use of small stripes allows for fine-grained load balancing between workers
with different performance characteristics. The permutation is cheap to compute
and does not require any precomputed mapping tables.

The presented approach also allows for hierarchical workload distribution.
For example, after first statically distributing the stripes over machines in a
network, each host can further (possibly adaptively) distribute its stripes over
its compute devices (see Fig.~\ref{Fig:schedflowchart}). 
Because any contiguous interval of stripes has a good uniform distribution in image
space, the host can simply split its stripe interval up into contiguous
subintervals for each compute device.
Since the devices share the same permutation as the hosts, the partial host
framebuffer can be obtained from the partial device framebuffers by simply concatenating
the device framebuffers. Only the master host has to scatter the partial host framebuffers
into the final framebuffer.

\subsubsection{Batch Rendering}\label{Sec:Batch}

While interactive scheduling is optimized for rapid progressive feedback, 
the batch scheduling mode is optimized to minimize the total time to convergence.
The batch mode still generates intermediate progressive images but not as regularly.
In interactive scheduling, device synchronization and imperfect load balancing cost performance.
Furthermore, as the framebuffer is partitioned over more devices, each device will have less opportunities for parallelism.
This limits the scalability of the interactive scheduling mode.

\begin{figure*}[ht]
  \centering
  \includegraphics[height=0.2235\linewidth]{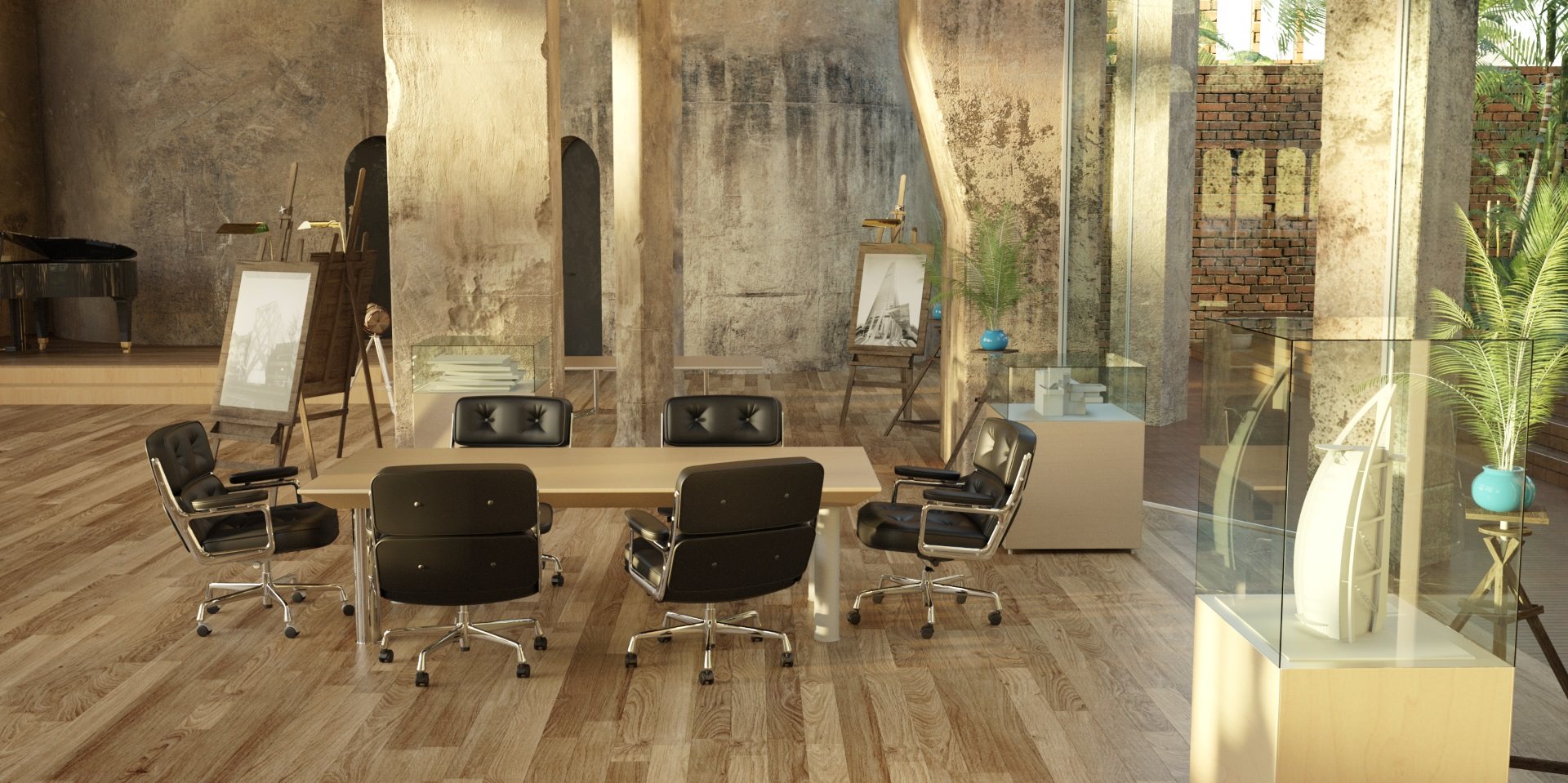} \hfill
  \includegraphics[height=0.2235\linewidth]{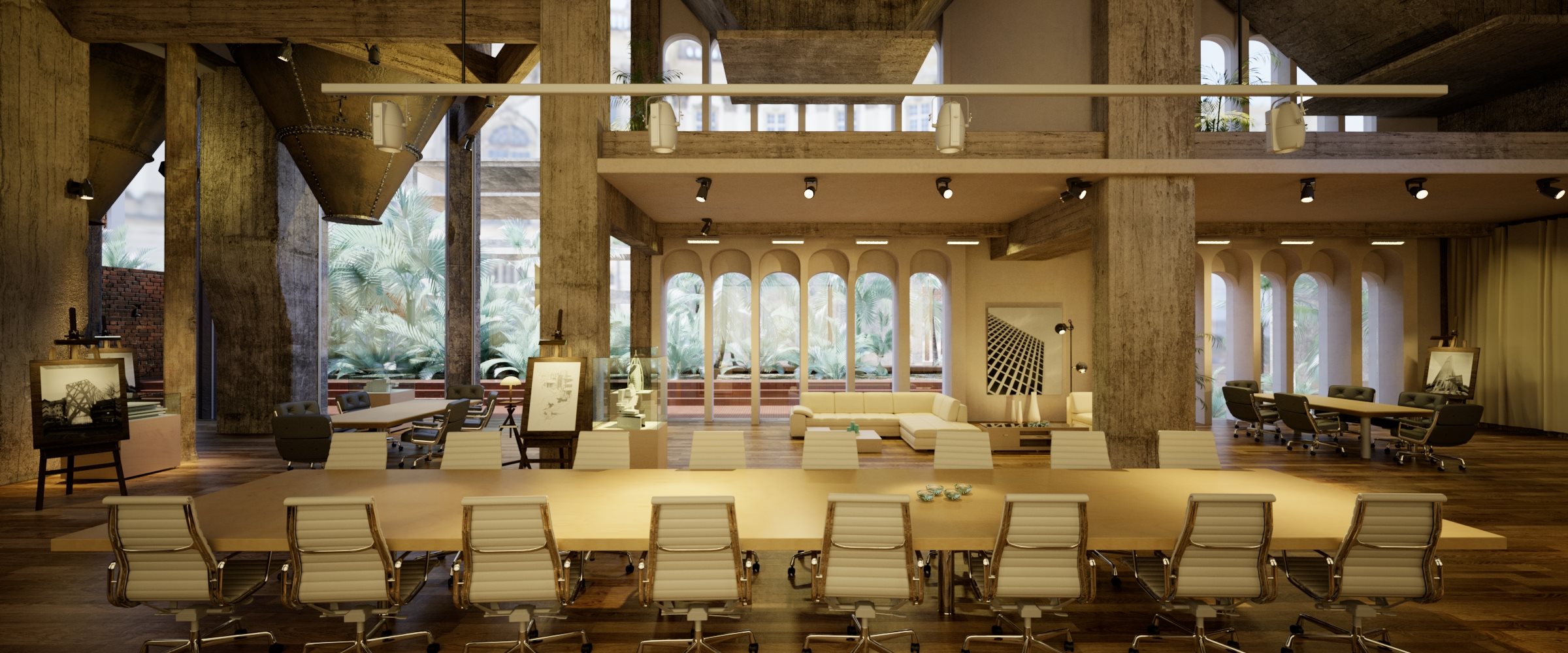}
  \caption{Barcelona Loft. The interior features lighting from both the environment and many small geometric light sources. \label{Fig:Interior}}
\end{figure*}

To eliminate these issues in batch scheduling, all devices render asynchronously (see Fig.~\ref{Fig:schedflowchart}).
Each device works on a local copy of the full framebuffer.
The low discrepancy sequences are partitioned in iterations of one sample per pixel. 
Sets of iterations are assigned dynamically to the devices.
This way a high per device workload is maintained, scaling to many devices.
To prevent congestion, the device framebuffer content is asynchronously merged into the host framebuffer in a best effort fashion.
If a device finishes an iteration set and another device is busy merging, the device will postpone merging and continue rendering its next assigned iteration set.
This mechanism could lead to starvation, possibly blocking some devices from ever merging,
and increase the potential of loss of progress in case of device failure.
Iray resolves this issue by also skipping merging if a device has less than half as many unmerged iterations as any other device on that host, allowing the other device to merge first.
Similarly, on a network level, all local progress is accumulated in the host framebuffer and is only periodically sent over and merged into the master framebuffer.
Iteration sets are assigned dynamically to hosts as needed, who further assign the iterations dynamically to the individual devices.
Similar to the interactive mode, if a device or host fails, Iray will continue rendering on the remaining devices.
Any iterations that were not merged yet are lost and reassigned for rendering to the remaining devices. 
As measured in Tab.~\ref{Tab:Scalability}, Iray exposes near linear performance scaling on a cluster of NVIDIA VCAs.

Device performance increases with iteration set size (see Tab.~\ref{Tab:Efficiency})
due to the increased opportunities for parallelism when the device generates multiple iterations at once.
However, if iteration sets are too big, load balancing will suffer from a tail effect when some devices are already done while others still have many iterations assigned.
Because a device works on an iteration set as a whole, iteration stealing among devices cannot be used to resolve this tail issue.
Instead the scheduler increases the assignment granularity as rendering completion approaches.
The remaining render time is estimated based on the observed rendering performance. The size of an iteration set assigned to any particular device is adjusted so that rendering of the iteration set
is expected to take no more than half the expected remaining render time. This eliminates most of the tail effect without the need for work stealing.

\begin{landscape}
	\begin{table}[t]
		\centering
		\begin{tabular}{| r | r | r | r | r | r | r | r | r | r | r | r | r |}
			\hline
			Iterations & \multicolumn{3}{c|}{Still Life, 1280x720 (Fig.~\ref{Fig:StillLife})} & \multicolumn{3}{c|}{Barcelona Loft, 1024x576 (Fig.~\ref{Fig:Interior})} & \multicolumn{3}{c|}{Lobby, 1920x1080 (Fig.~\ref{Fig:LPE:recolor})} & \multicolumn{3}{c|}{BMW, 1280x720 (Fig.~\ref{Fig:Car2})} \\
			per frame  & \multicolumn{3}{l|}{} & \multicolumn{3}{l|}{} & \multicolumn{3}{l|}{} & \multicolumn{3}{l|}{} \\
			\hline
			& wf. & mk. & speedup & wf. & mk. & speedup & wf. & mk. & speedup & wf. & mk. & speedup \\
			\hline
			$\frac{1}{64}$ & 0.67 & 1.29 &  51.61\% & 0.48 & 0.82 &  58.25\% & 0.25 & 0.24 & 104.93\% & 0.70  &  1.63 &  43.03\%\\
			$\frac{1}{16}$ & 1.79 & 2.55 &  70.13\% & 1.21 & 1.51 &  79.88\% & 0.52 & 0.30 & 173.36\% & 2.09  &  4.43 &  47.05\%\\
			$\frac{1}{4}$  & 4.03 & 3.38 & 119.19\% & 2.32 & 1.82 & 127.52\% & 0.74 & 0.32 & 233.47\% & 5.57  &  8.76 &  63.61\%\\
			          1.00 & 6.51 & 3.72 & 175.08\% & 3.27 & 1.96 & 166.72\% & 0.84 & 0.32 & 261.89\% & 12.30 & 12.02 & 102.40\%\\
			          4.00 & 7.57 & 3.77 & 200.50\% & 3.58 & 1.98 & 180.73\% & 0.92 & 0.32 & 289.57\% & 18.94 & 13.13 & 144.23\%\\
			         16.00 & 7.81 & 3.77 & 206.92\% & 3.69 & 1.98 & 186.30\% & 0.95 & 0.32 & 297.89\% & 21.79 & 13.40 & 162.63\%\\
			         64.00 & 7.88 & 3.77 & 208.96\% & 3.69 & 1.98 & 186.33\% & 0.96 & 0.32 & 299.92\% & 23.12 & 13.40 & 172.51\%\\ \hline
		\end{tabular}
		\caption{Iterations per second on a single NVIDIA Quadro P6000 GPU as function of the iterations per frame, for 
			wavefront (wf.) and megakernel (mk.) approaches, and the relative speedup of wavefront over megakernel. 
			Comparable fractional iterations per frame where measured as a single iteration at reduced resolution.
			Note that image resolutions differ, so iterations per second are not directly comparable between the four scenes.
			The efficiency of the wavefront approach using just-in-time queues consistently increases with the number of
			iterations per frame, surpassing the performance of the megakernel approach. \label{Tab:Efficiency}}
	\end{table}
	
	\begin{table}[t]
		\centering
		\begin{tabular}{| r | r | r | r | r | r | r | r | r |}
			\hline
			Hosts & \multicolumn{2}{c|}{Still Life (Fig.~\ref{Fig:StillLife})} & \multicolumn{2}{c|}{Barcelona Loft (Fig.~\ref{Fig:Interior})
			} & \multicolumn{2}{c|}{Lobby (Fig.~\ref{Fig:LPE:recolor})} & \multicolumn{2}{c|}{BMW (Fig.~\ref{Fig:Car2})} \\
			\hline
			& Iterations/300s & Efficiency & Iterations/300s & Efficiency & Iterations/300s & Efficiency & Iterations/300s & Efficiency \\
			\hline
			1  & 30787  &100\% & 6571   & 100\%& 4259  &100\%& 91572  & 100\%\\
			2 & 60888  & 99\% & 12602   &  96\%& 8420 &99\%& 183213  &  100\%\\ 
			4 & 121132 & 98\% & 25294   &  96\%& 16739 &98\%& 365418 &  100\%\\
			8 & 240185 & 98\% & 51369  &  98\%& 33178 & 97\%& 723195 &  99\%\\
			\hline
		\end{tabular}
		\caption{Very good scalability across compute cluster configurations:
			For a fixed interval of time (300 seconds), the number of completed iterations has been counted for an increasing
			number of hosts, where each host rendered on 8 NVIDIA Quadro M6000 GPUs and a Xeon E5 CPU. The efficiency of
			scaling is the ratio of the iterations in the system divided by the number
			of hosts and the iterations on one host.
			\label{Tab:Scalability}}
	\end{table}
\end{landscape}

\section{Rendering Workflows} \label{Sec:Workflows}

Many traditional production rendering systems rely on a large
amount of user input to achieve convincing lighting. 
Light transport simulation allows workflows to shift from manually 
placed point light sources with associated shadow maps to the 
push-button lighting of a scene using physically-based entities.
Since the simulated light behaves just like light in the real world,
many effects that are tedious to create with traditional systems simply happen automatically.

A light transport simulation system that is able to compute
results that are sufficiently accurate to replace measurements
of the real world is called predictive. Such systems enable designers
and architects to save many iterations of building physical prototypes.
In combination with progressive image feedback
this allows for much faster turnaround times when creating and changing
virtual scenes and looks.

\subsection{Light Path Expressions (LPEs)} \label{Sec:LPE}

Practically all image synthesis workflows, whether they focus on realism or artistic expression, involve some degree of post 
production work.
Such processes typically involve the composition of partial images (referred to as passes or layers) to create the final result.
The types of these layers can be roughly partitioned into layers which are based on light transport simulation, and auxiliary layers 
which contain normal, position, depth, or other attribute information. While similar in spirit to Arbitrary Output Values (AOVs) in other renderers, LPEs provide a more flexible and expressive solution in situations where the desired result is generated by light transport simulation. As such, they are a natural fit to our architecture, as our implementation makes no assumptions concerning the desired content of the layers and at the same time provides full control to the user.

LPEs are regular expressions that describe the interactions of a light transport path with the scene \cite{Heckbert:1990:ART}.
Interactions can be distinguished by a combination of type (emission, reflection, refraction, or volume scattering), 
mode of scattering (diffuse, glossy, or perfectly specular), or name of scene element or material component.

LPEs operate like a filter on path space, so that the contribution of each generated path is only written to those layers whose 
associated expression matches. For example, a user could specify an LPE that contains only glossy bounces from one particular light source.

Because the filter is applied to each generated path, it affects all simulated effects, including dispersion, motion blur, and depth of 
field. If, during the construction of a path, it becomes clear that none of the specified LPEs will match, the sample is terminated prematurely.

Common issues with separating objects in pixels with multiple visible objects are avoided for the same reason, which drastically improves anti-aliasing in the composited result.
This obviates the needed for alpha channels in most composition cases.
Nonetheless, the alpha channel of each result layer may also be controlled by an additional LPE. 

In particular, no modifications of the scene are necessary, besides assigning names as needed.
Note that AOVs may allow for capturing non-physical effects which are not covered by Iray and its LPEs.

\begin{figure}
\includegraphics[width=0.3215\linewidth]{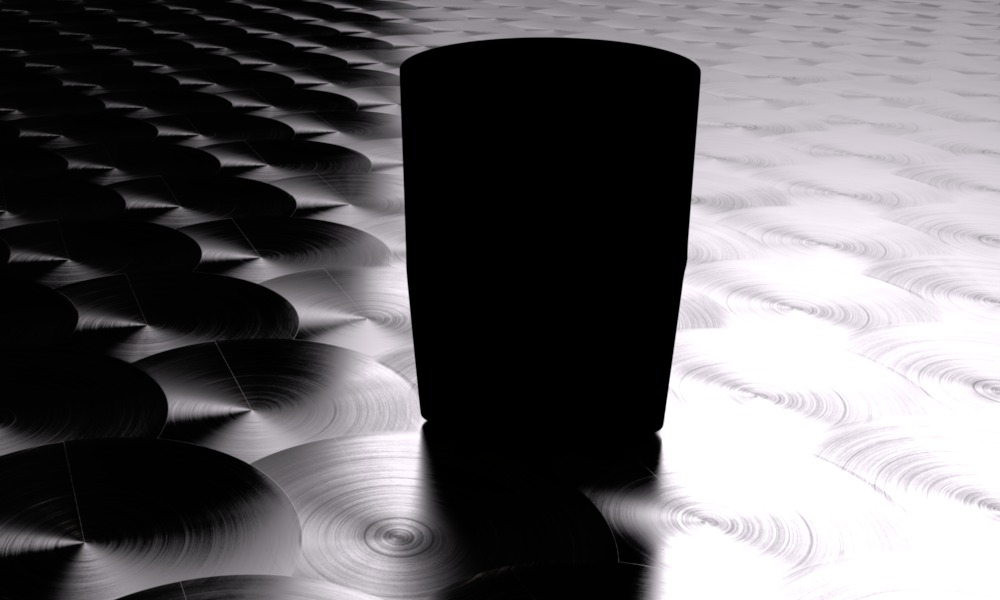} \hfill
\includegraphics[width=0.3215\linewidth]{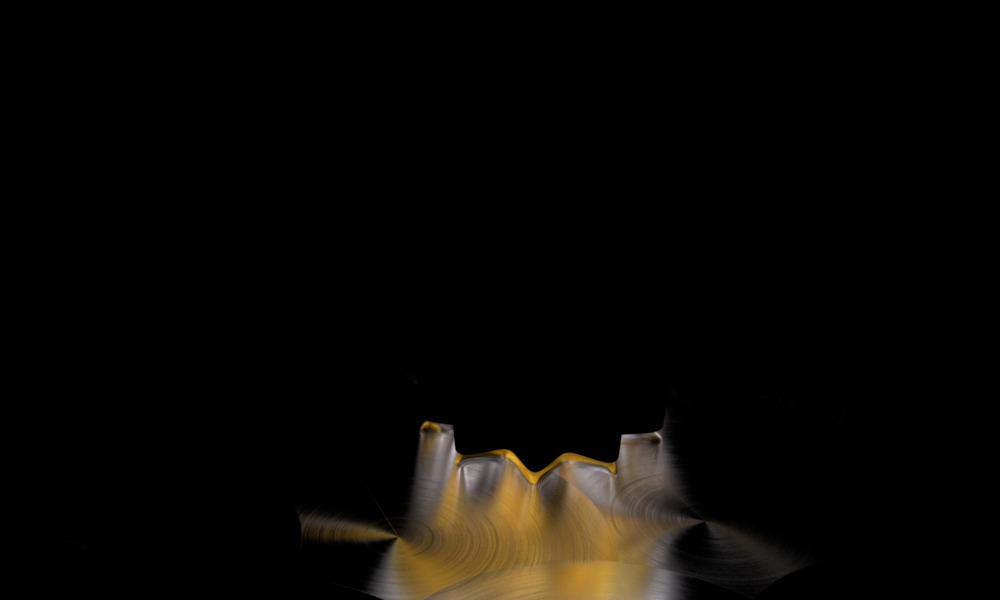} \hfill
\includegraphics[width=0.3215\linewidth]{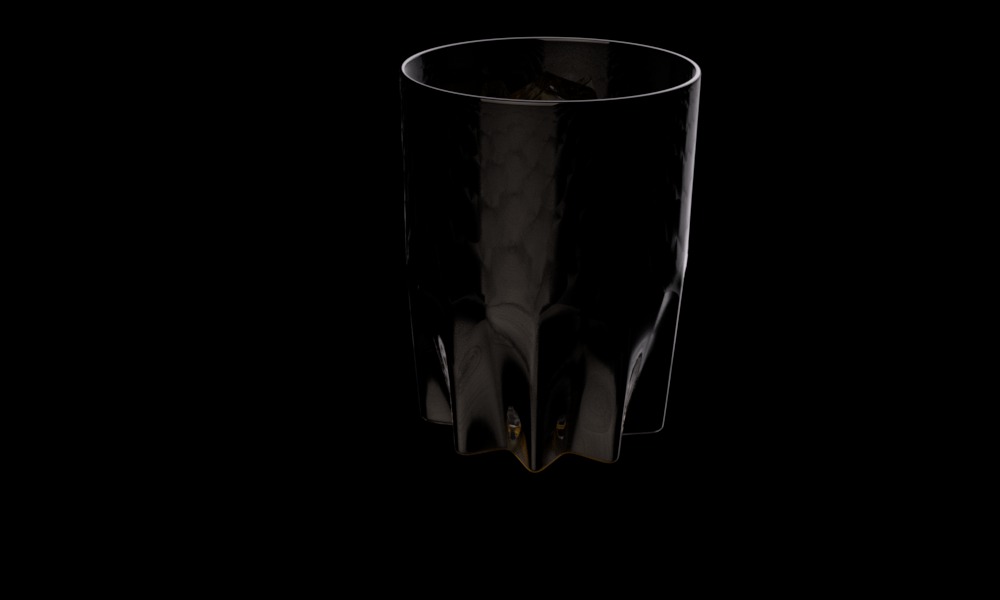} \\[2mm]
\includegraphics[width=0.3215\linewidth]{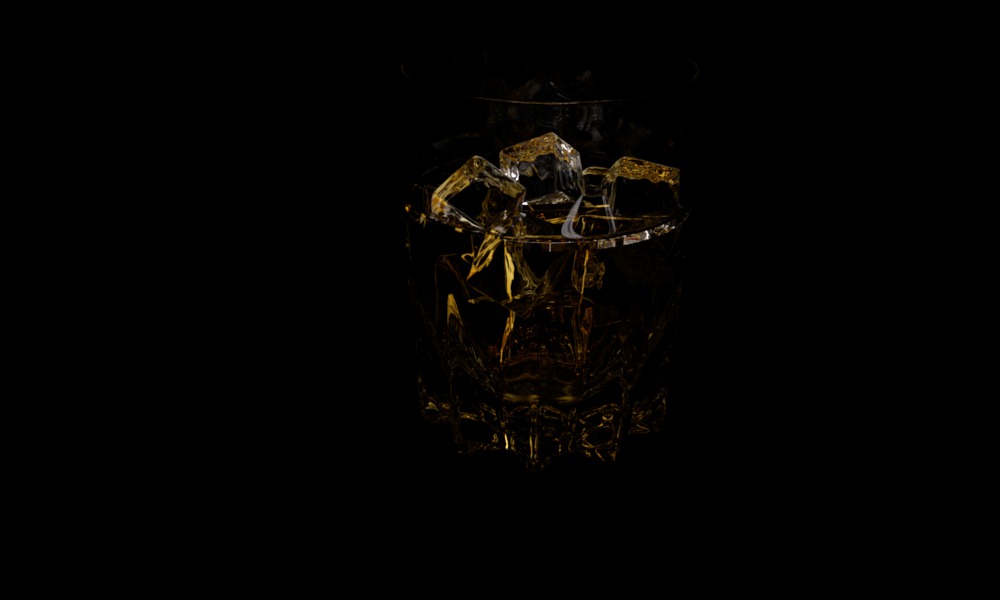} \hfill
\includegraphics[width=0.3215\linewidth]{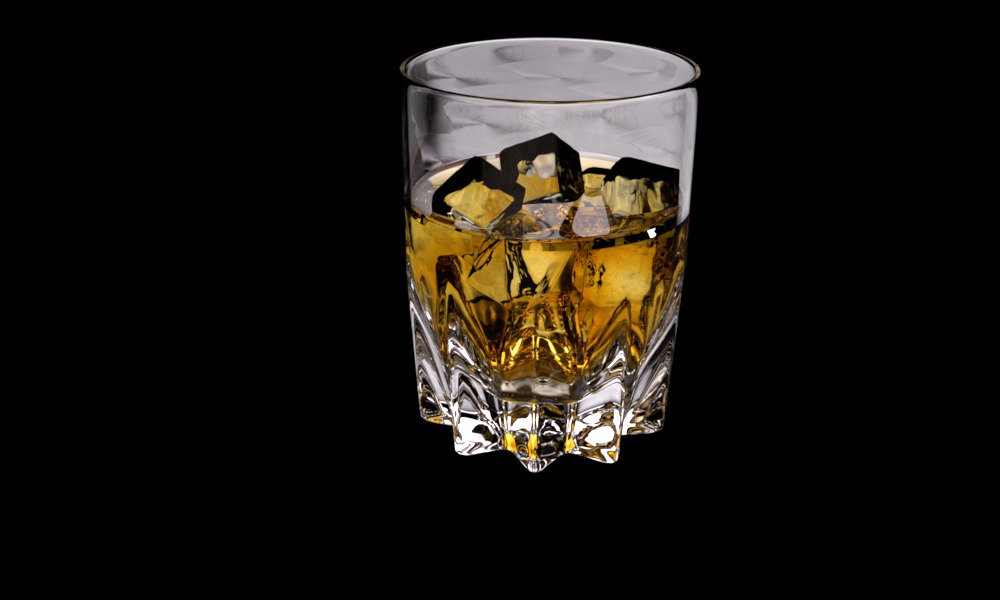} \hfill
\includegraphics[width=0.3215\linewidth]{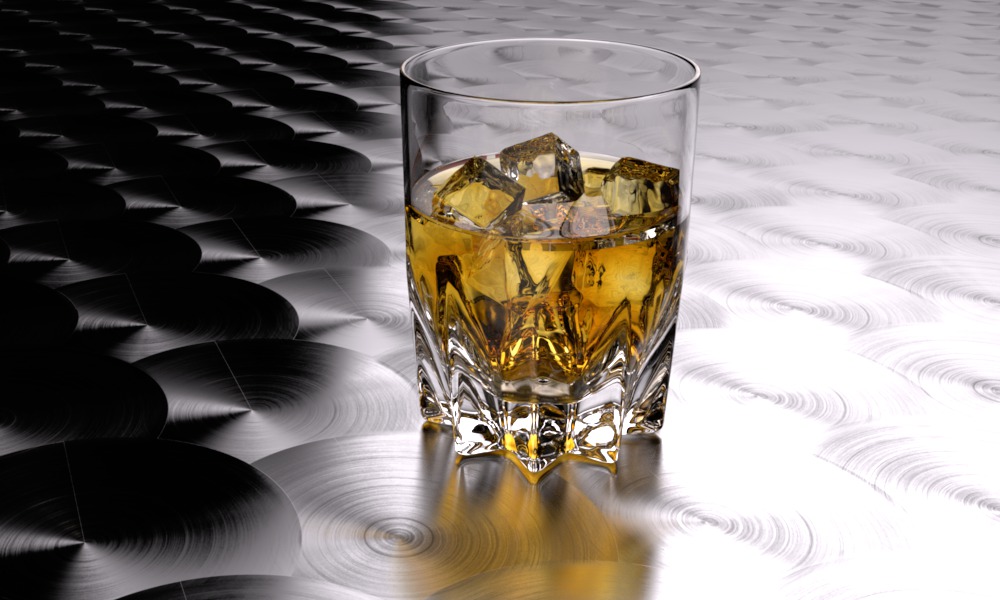}
    \caption{An example decomposition of light transport using LPEs. \label{Fig:LPE:elements}
        From top left to bottom right:
        Light falling onto the ground without first passing through the glass.
        Caustics cast by the glass.
        Specular reflections on the glass.
        Specular reflections on the ice cubes.
        All remaining interactions.
        The full solution, i.e.\ the sum of the previous images.}
\end{figure}

\begin{figure}
    \centering
    \includegraphics[width=0.3215\linewidth]{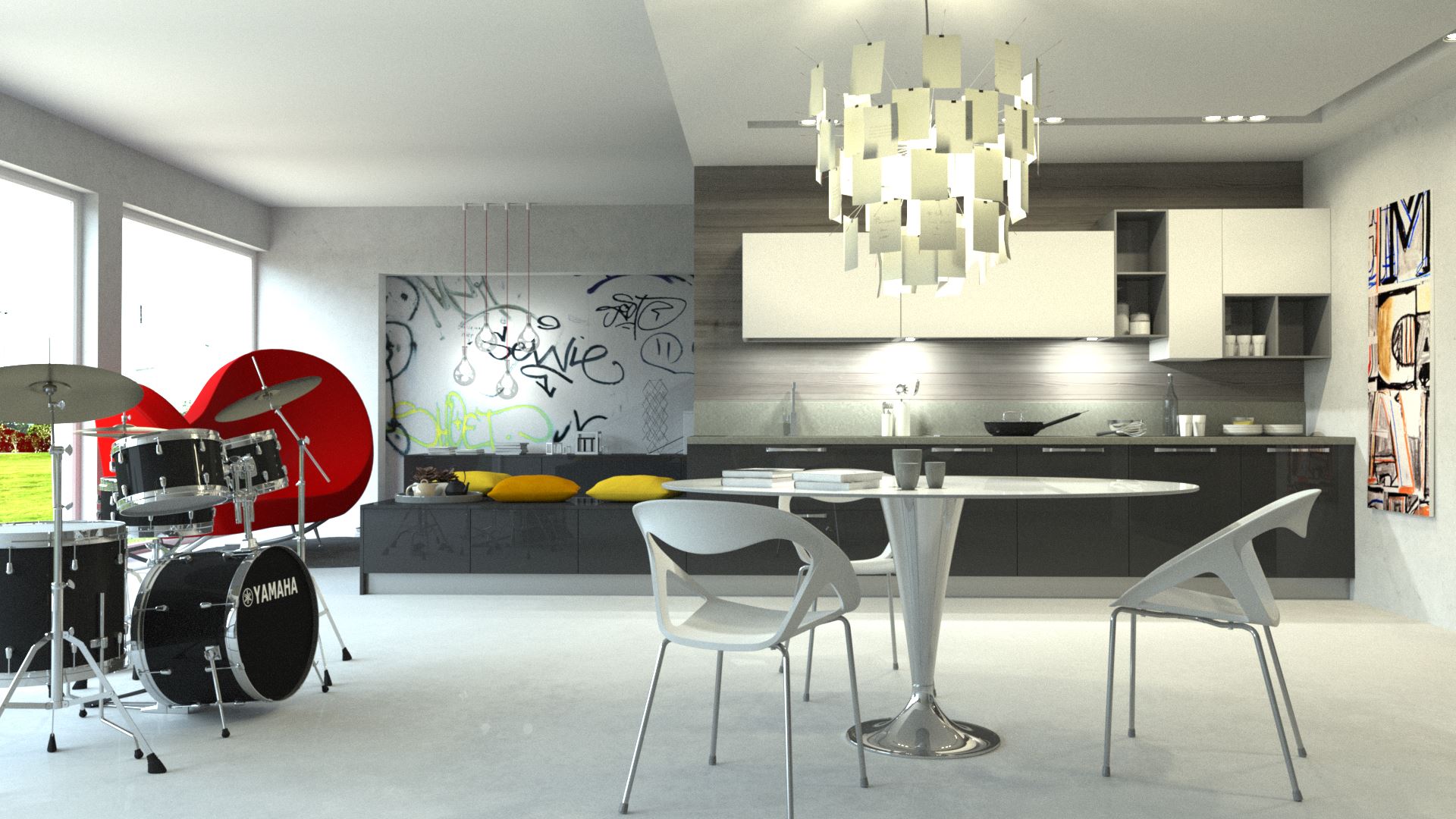} \hfill
    \includegraphics[width=0.3215\linewidth]{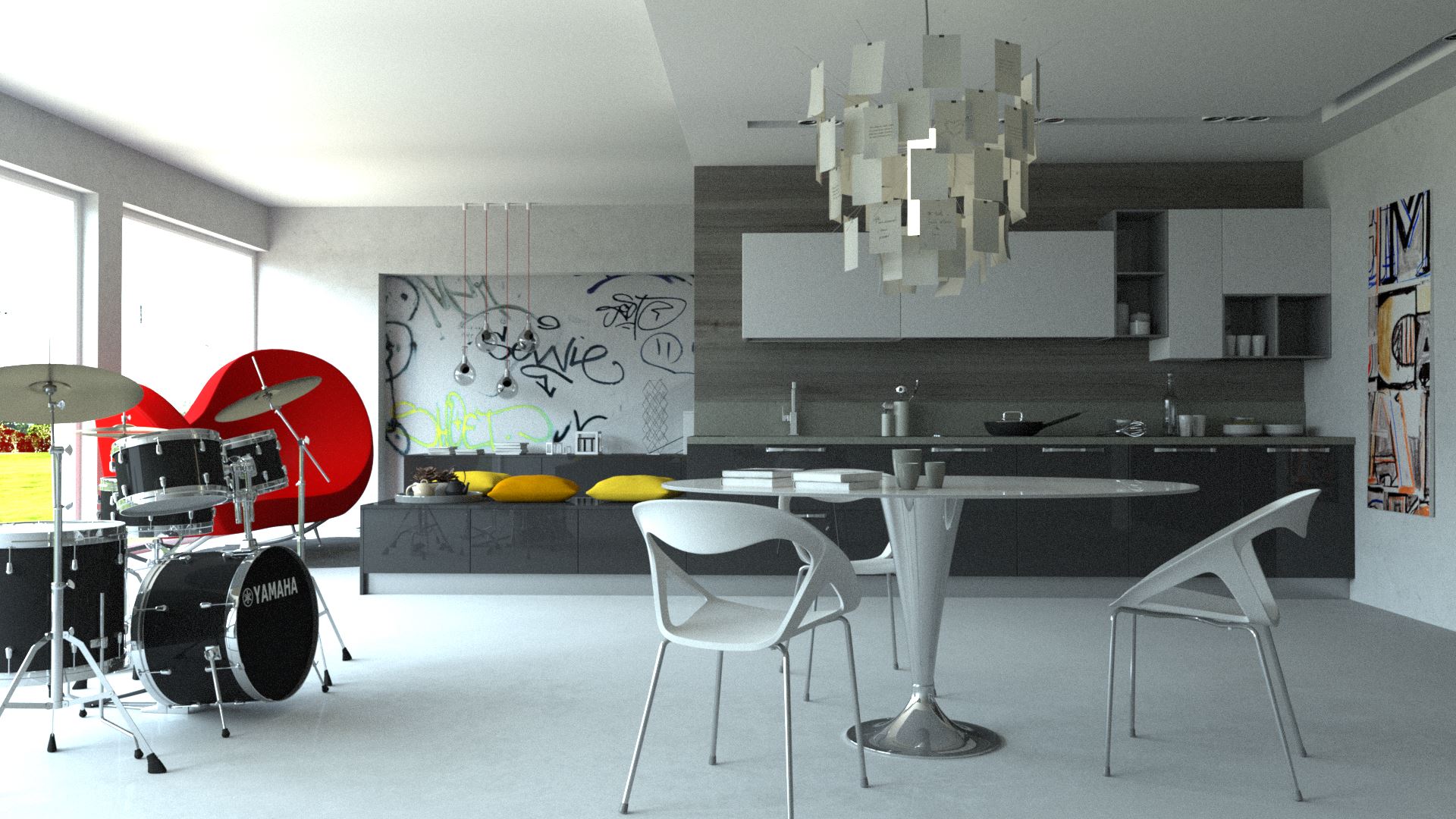} \hfill
    \includegraphics[width=0.3215\linewidth]{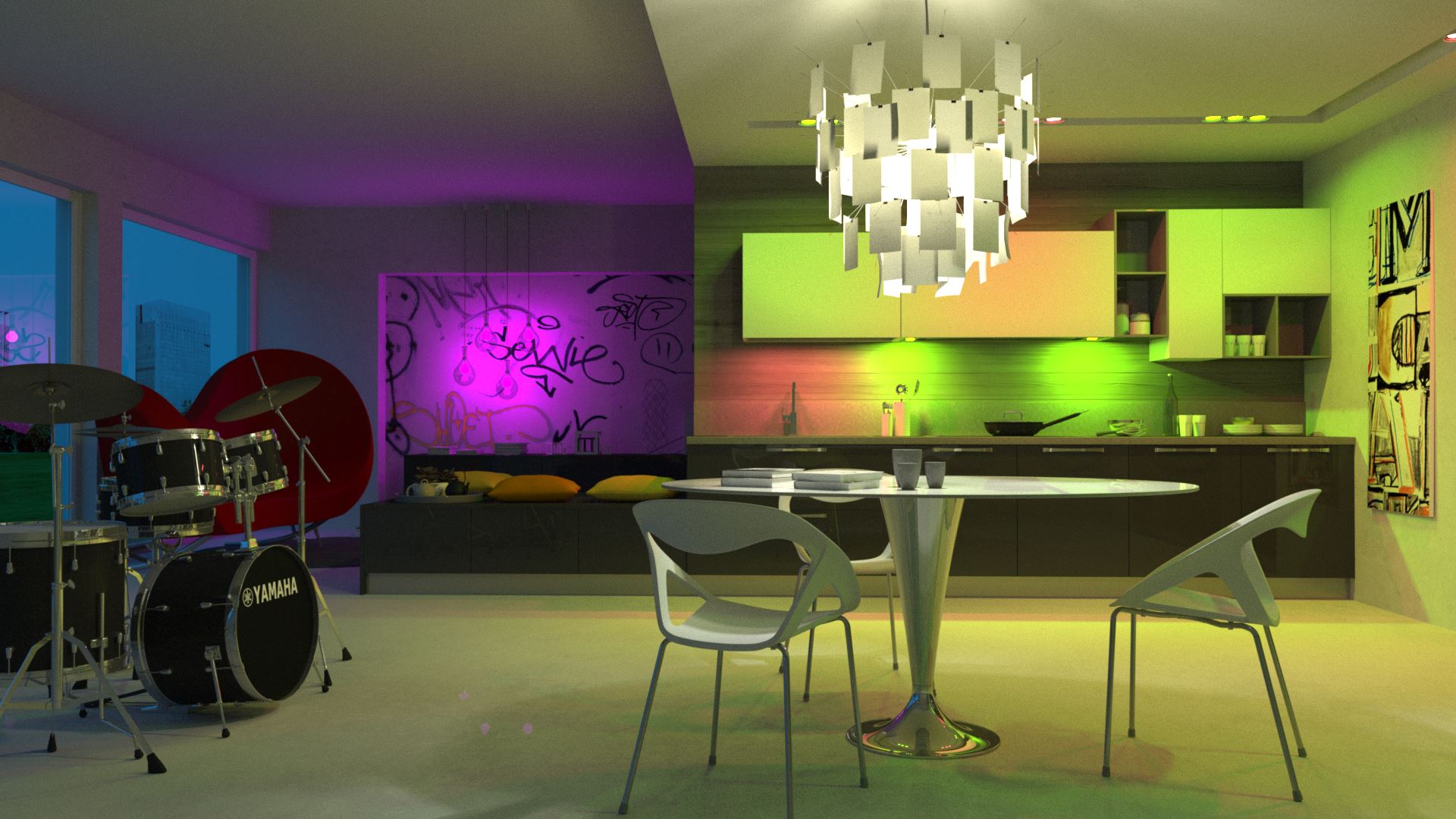}
    \caption{Relighting a scene during post processing. 
        All images are generated from the same set of LPE images. 
        The images match exactly the result that would have been obtained after modifying the scene and re-rendering.
        \label{Fig:LPE:relight}}
\end{figure}

\begin{figure}
        \centering
        \includegraphics[width=0.3215\linewidth]{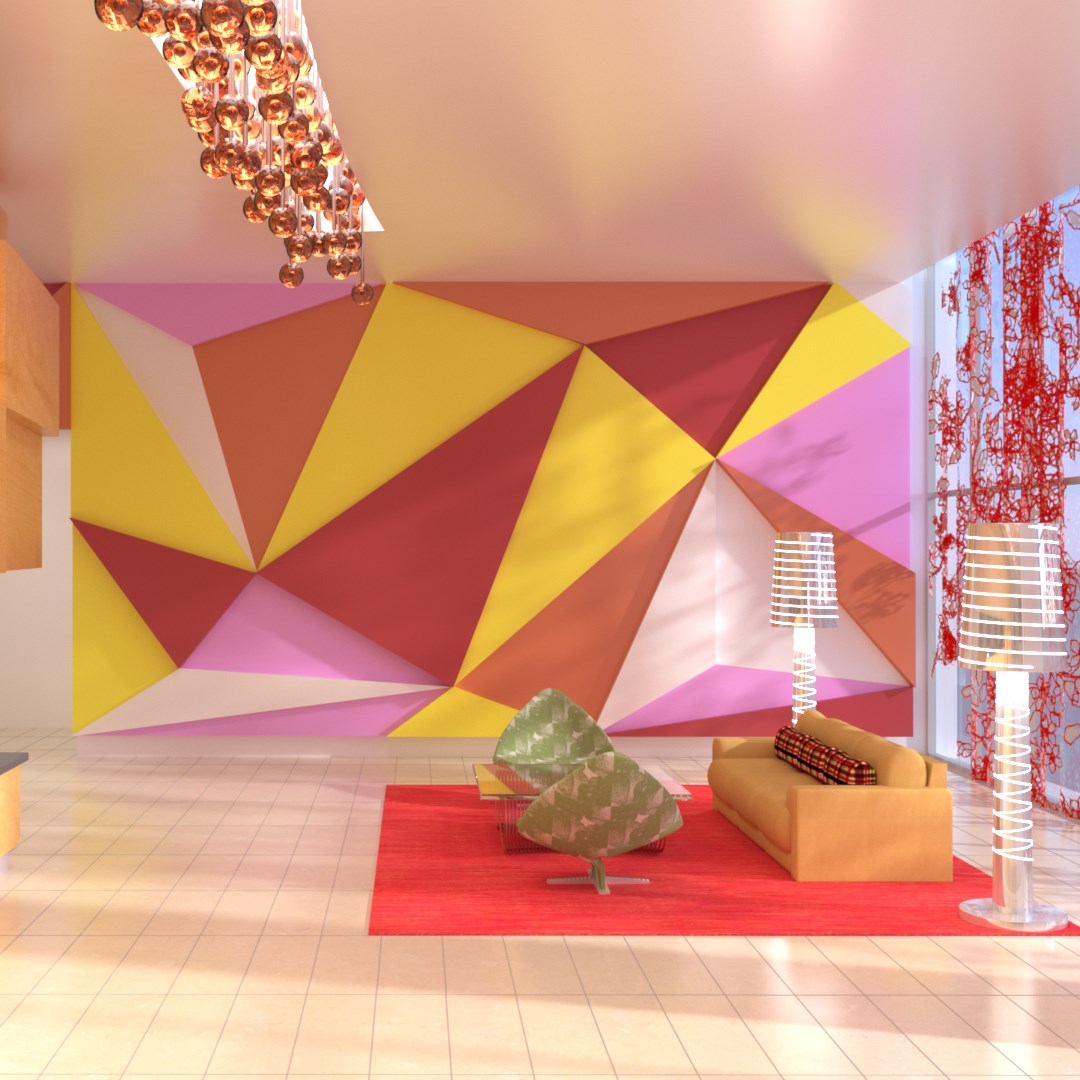} \hfill
        \includegraphics[width=0.3215\linewidth]{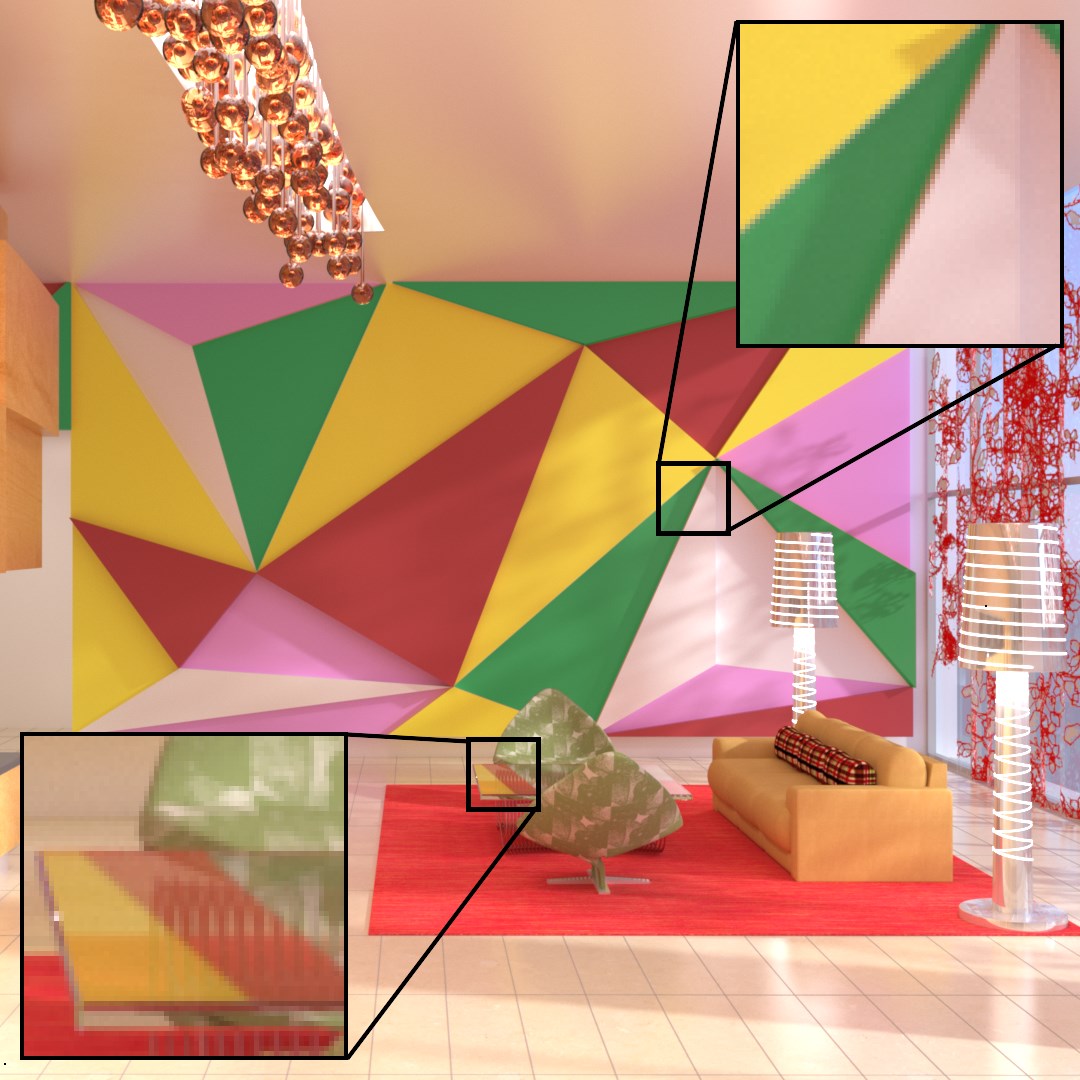} \hfill
        \includegraphics[width=0.3215\linewidth]{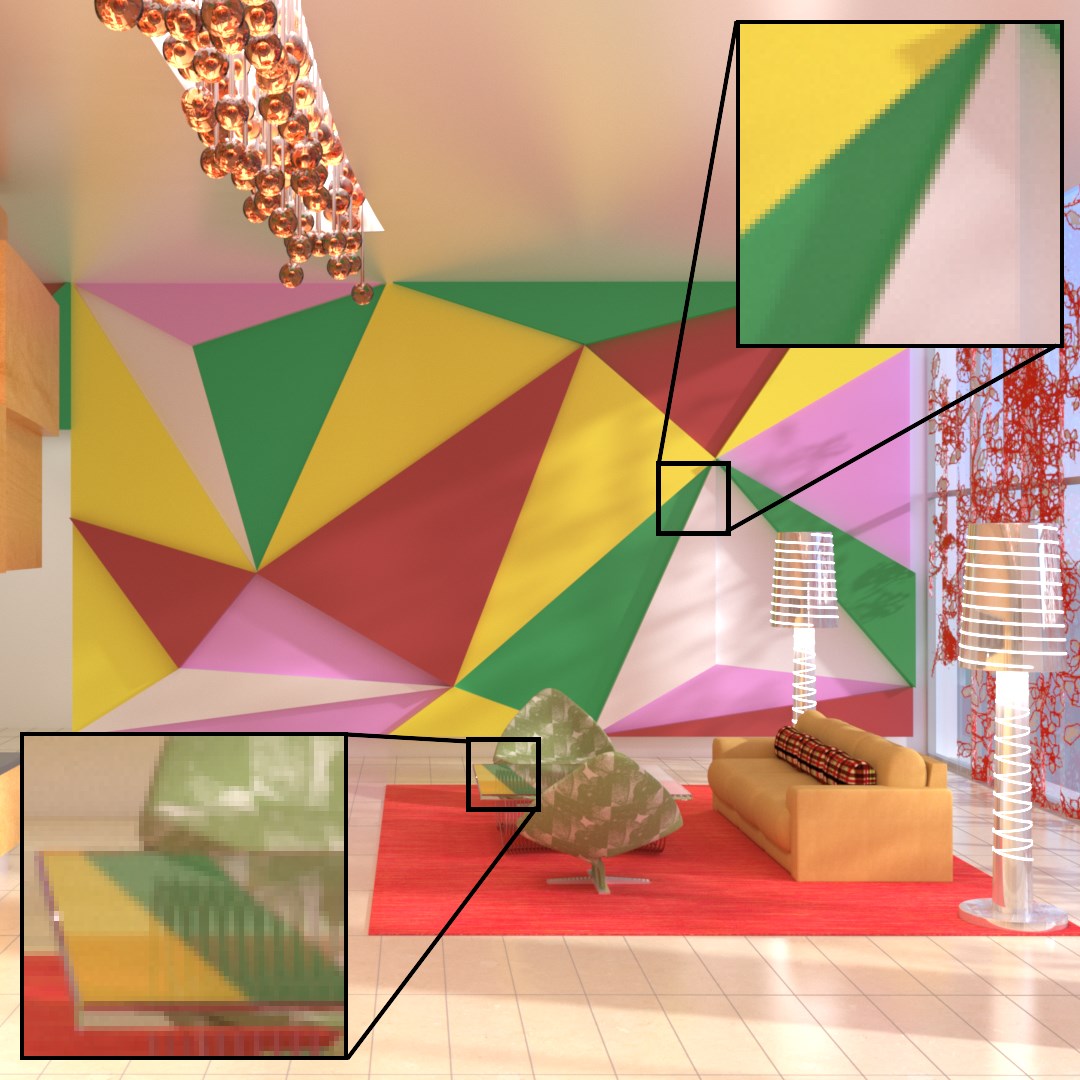}
        \caption{Lobby. Recoloring a wall during post processing with LPEs. From left to right: Original beauty output. Manual recoloring using a traditional mask-based approach. Recoloring using LPEs.
        Note the appropriately updated secondary effects (reflection on the glass table, inset on the lower left) and proper edge handling (inset on the upper right) when using LPEs instead of manual pixel masking.
        \label{Fig:LPE:recolor}}
\end{figure}

The system makes it trivial to achieve effects like removing or attenuating unwanted highlights (see Fig.~\ref{Fig:LPE:elements}), or adding 
a bloom effect only to visible specular elements.
However, more interesting things become possible:
For instance, the color and brightness of any number of light sources can be changed instantly in post production, without having 
to re-render anything, essentially relighting the scene without having to re-render it while still retaining all global illumination (Fig.~\ref{Fig:LPE:relight}). 
Similarly, it is possible to change the color of an object after the rendering, automatically accounting for secondary effects like color bleeding or 
reflections, as illustrated in Fig.~\ref{Fig:LPE:recolor}. 

Results obtained by post-modifying the color or intensity of light sources in a linear color space retain the physical accuracy of the original rendering, 
which is especially important when working with irradiance probes (see Sec.~\ref{Sec:probes}). %
However, LPEs allow for tweaks that may exceed the boundaries of physical correctness by far.
While this is often desirable in artistic workflows, care should be exercised in predictive rendering scenarios.

For example, a common approach to the recoloring illustrated in Fig.~\ref{Fig:LPE:recolor} is applying a white material to a given object and separating all paths that interact with it.
Multiplying the result with the desired color and adding all remaining paths yields convincing images.
However, this result is not correct because it does not properly apply stronger tinting to paths that interact with the object more than once.
In most realistic scenarios the error of these approaches can be reduced to acceptable levels by using different layers for paths that interact with the object once, twice, and up to a low number of times, which is straightforward with LPEs.
Note that this type of inaccuracy affects any compositing in a global illumination setting and is not limited to or caused by LPEs.

\subsection{Alternate Outputs}\label{Sec:probes}

While the main focus of Iray is the generation of 2D (or stereo) images, the system also supports alternate output modes.
Rather than tracing rays from the camera, the user may choose to place a number of probe points anywhere in the scene. 
Iray then measures the irradiance impinging on those points from the hemisphere defined by a normal vector provided with each point.
The hemisphere is sampled progressively just like pixels are in the image generation mode.

This irradiance probe rendering mode may be used e.g.\ to determine whether workspaces are lit appropriately by generating a grid of probes at desktop height and simulating results at various times of the day.
This was used during the planning of NVIDIA's new headquarter building in Santa Clara in parallel with conventional tools used by lighting consultants~\cite{gensler:GTC2016}.
Results obtained during this process matched conventional tools in overall quantities~\cite{ciereport} but yielded more detailed results.
Using these findings, it was possible to optimize artificial lighting and placement of windows already during the design stage.

In combination with LPEs, more detailed studies like comparing direct to indirect irradiance, 
daylight versus artificial lighting contributions, or glare, are simple to perform.

Iray may also be used to generate 4D light fields that allow for realtime exploration of photoreal environments in virtual reality applications.

\subsection{Matte Objects} \label{Sec:matte}

One very important aspect of digital content creation is merging computer generated (CG) objects 
and real-world backplate and/or environment photographs \cite{RenderSynthObj}. This section discusses how Iray meets this
practical requirement without sacrificing consistency of the light transport simulation.

Backplate photographs are usually employed as a high-quality background in conjunction with a spherical environment 
that is used to approximate all reflections and lighting from the real-world surroundings.
Alternatively (and ideally), a single high-quality spherical environment may be used to fill both roles.

In this context, matte objects are placeholder geometry that is created as a stand-in for selected objects in the backplate 
scene, like buildings or streets that are visible in a backplate photograph. They are used to simulate interactions between CG objects and the backplate and are crucial to making the CG objects 
appear as though they were part of the background scene.

\begin{figure}
  \centering
  \includegraphics[width=\linewidth]{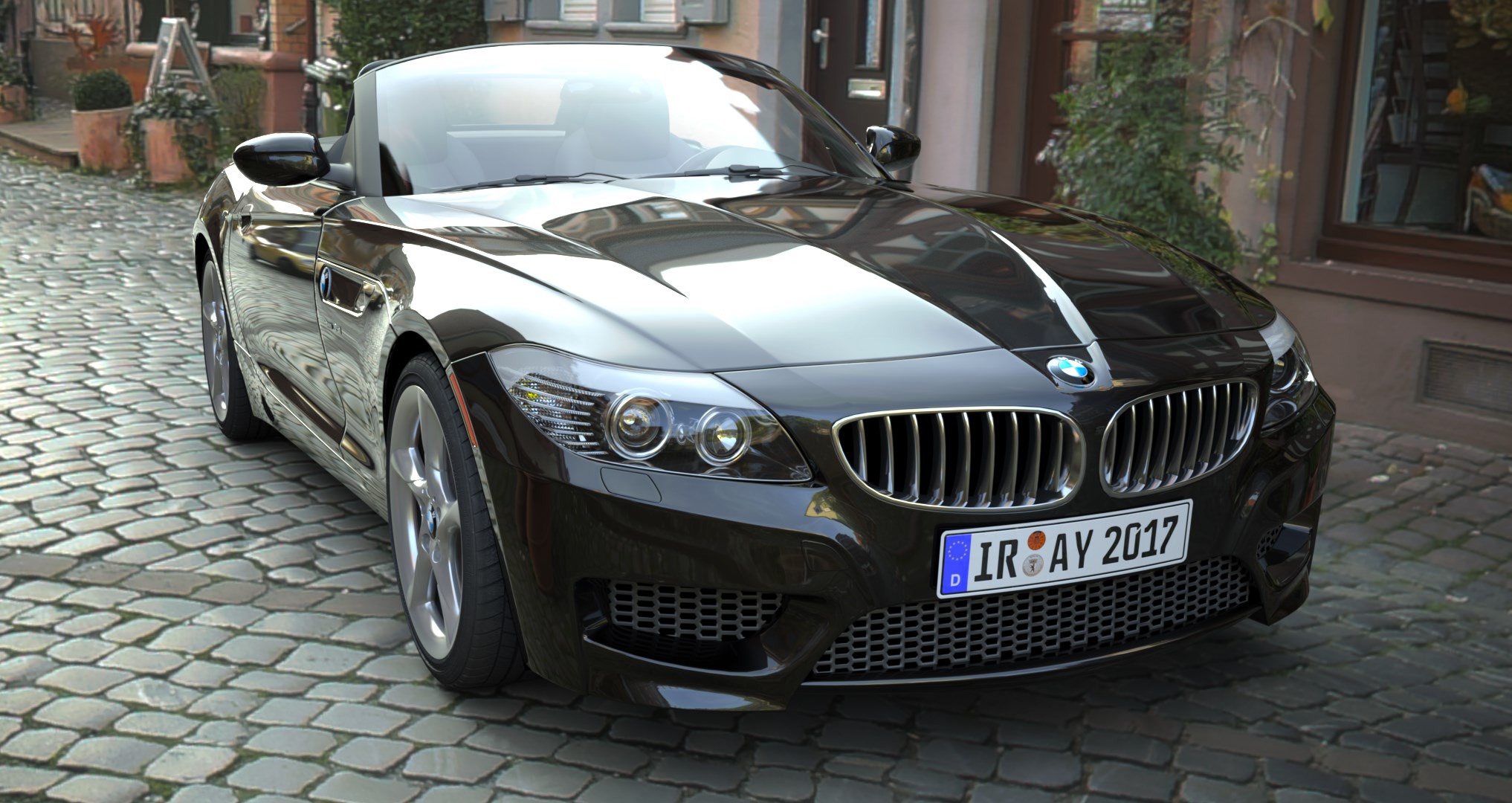}   \caption{BMW. The invisible ground plane is flagged as matte object and can receive shadows 
based on the actual light sources to tone down existing radiance in the backplate. In addition, secondary effects seen on the computer generated car, such as reflections and refractions, also interact seamlessly with both the matte object and backplate. \label{Fig:Car2}}
\end{figure}

Traditionally, these interactions are computed separately from the main rendering process and are then composited in a final 
post-processing step. In contrast, our system applies physically-based rendering paradigms to matte object rendering while keeping 
the feature as simple to use as possible. Consequently, matte objects are handled directly within the light transport simulation and do not jeopardize the correctness of the simulation or the progressive nature of the rendering.
This enables accurate secondary effects of matte objects, while still retaining pixel-perfect reproduction of supplied LDR backplates in the final image, as would be the case with manual compositing.

In order to construct a scene with matte objects, the user first creates geometry that will act as a stand-in for 
objects in the backplate (a simple example being a flat ground plane to represent a street). 
This geometry is then associated with materials just like standard CG objects.
The directly visible appearance of matte objects will remain determined by the backplate photograph, but additional interactions
like reflections of CG objects in matte objects will be simulated based on the provided material and thus can be treated as any normal interaction.
Therefore, the better these materials match the materials in the backplate, the better the quality of the rendering will be. 
Finally, CG objects and additional light sources are added to the scene as usual.

Since the concept of shadows is non-existent in the simulation, but needed to modulate already existing radiance in the backplate, 
our system generates artificial shadows on matte objects by sampling light sources as described 
in Sec.~\ref{Sec:LightSources}, which delivers plausible shadows in most common cases (see Fig.~\ref{Fig:Car2}).
To give artists additional flexibility, we further allow for tweaking the intensity of the shadows. All the remaining interactions of matte objects and CG objects 
are generated directly by the actual simulation.

\begin{figure}
  \centering
  \includegraphics[width=0.4915\linewidth,trim={0 0 0 2cm},clip]{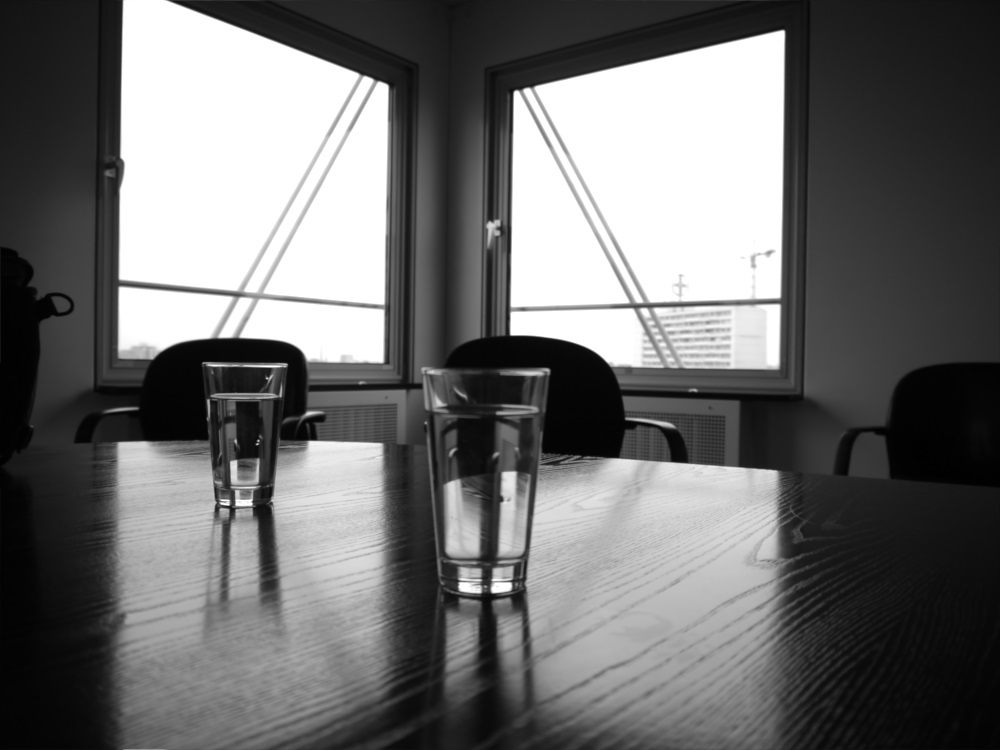} \hfill
  \includegraphics[width=0.4915\linewidth,trim={0 0 0 2cm},clip]{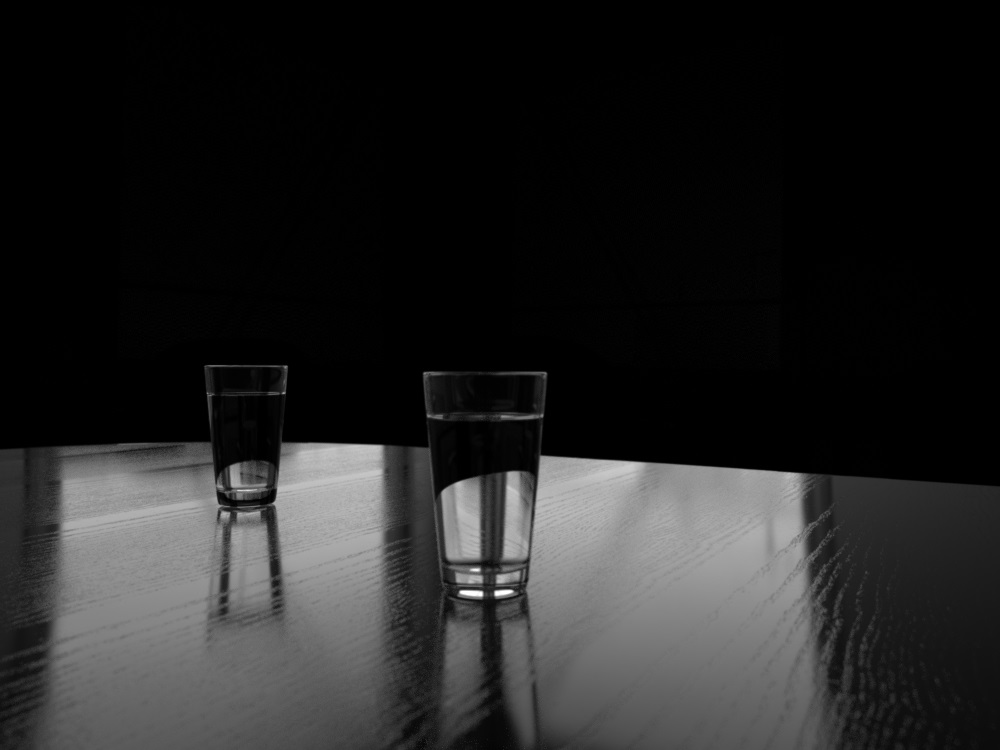}\\\vspace{2mm}
  \includegraphics[width=0.4915\linewidth,trim={0 0 0 2cm},clip]{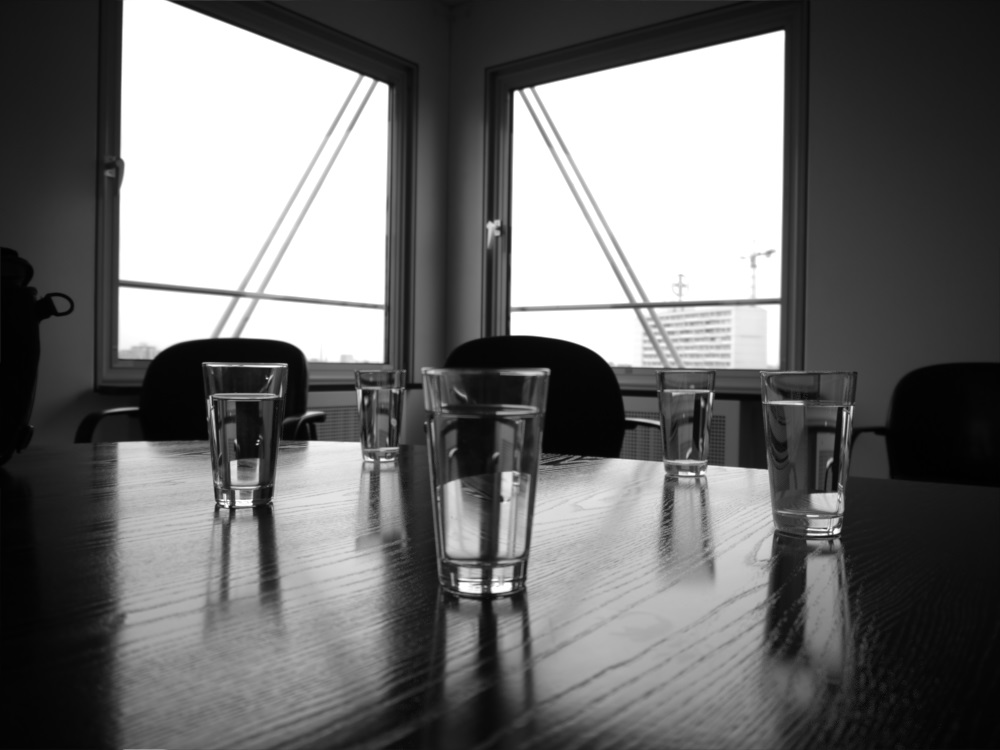} \hfill
  \includegraphics[width=0.4915\linewidth,trim={0 0 0 2cm},clip]{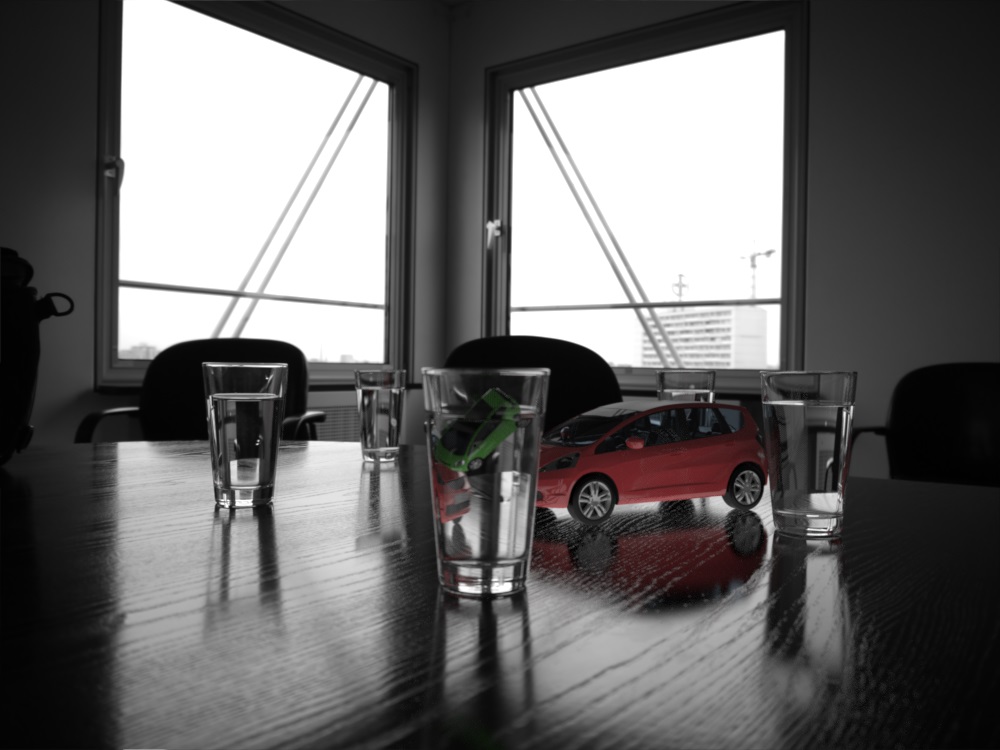}
  \caption{Top left: Original, raw photograph. Top right: The table and the glasses are re-created as CG objects with roughly matching materials and environment lighting. Bottom left: The table and glasses are now flagged as matte objects, the photograph is set as the backplate and some additional CG glasses are added to the scene. Bottom right: Two additional CG cars are added to the scene. 
  Note the interaction of CG objects with both CG and matte objects without the need for any further tweaking. \label{Fig:Matte}}
\end{figure}

Note that after intersecting a matte object, some of the collected light contributions will be dependent on future interactions of the path with the scene,
so an additional temporary storage per path is necessary. This is then used to composite the intermediate results
along with the shadow modulation before merging the final result to the framebuffer.
One example is the need to avoid multiple contributions of the backplate and environment on matte object interactions, as these are already part of the photograph itself (see Fig.~\ref{Fig:Matte}), which does not apply to the regular CG lights that might also be part of the same scene.
Other, more complex, examples are the interaction between different matte objects or the lighting received from matte emissive objects which includes shadows cast from matte objects onto other matte objects.

\subsection{Decals}

A common request for scene modeling is the possibility to add a variety of decals, like advertisement stickers or detail textures, on top of an object's material.
In the simplest case, where merely the color of an object is to be changed, this can typically be implemented within the texture blending mechanisms of the renderer.
If in addition also material properties are supposed to be different for the decal, this could (to some extend) still be realized using the layering facilities of the BSDF model.
However, such a workflow can easily yield very complex materials and their general usability seriously suffers, along with severe performance implications, which was 
the main reason to implement native decal support into the Iray core itself.

We made the workflow using decals simple and orthogonal to the material model in Iray, by creating a dedicated scene element driven by the following goals:
\begin{itemize}
\item
  The number of decals can be large.
\item
  One decal can cover multiple objects.
\item
  Decals can be placed upon decals.
\item
  Decals have their own (flexible) MDL material, including ``transparent'' decals using transmitting BSDFs.
\end{itemize}
Our implementation is in fact inspired by a na\"{\i}ve approach to creating decals: The use of actual geometry representing the decal, slightly offset from the base object.
Instead of real geometry (and inviting associated ray tracing precision issues), we merely pretend that it is present.
More precisely, in the rendering core we imagine decals as thin layers of virtual geometry that are simulated just as regular thin-walled geometry would be if separated by a very small air gap.

\begin{figure}
    \centering
    \includegraphics[width=\linewidth]{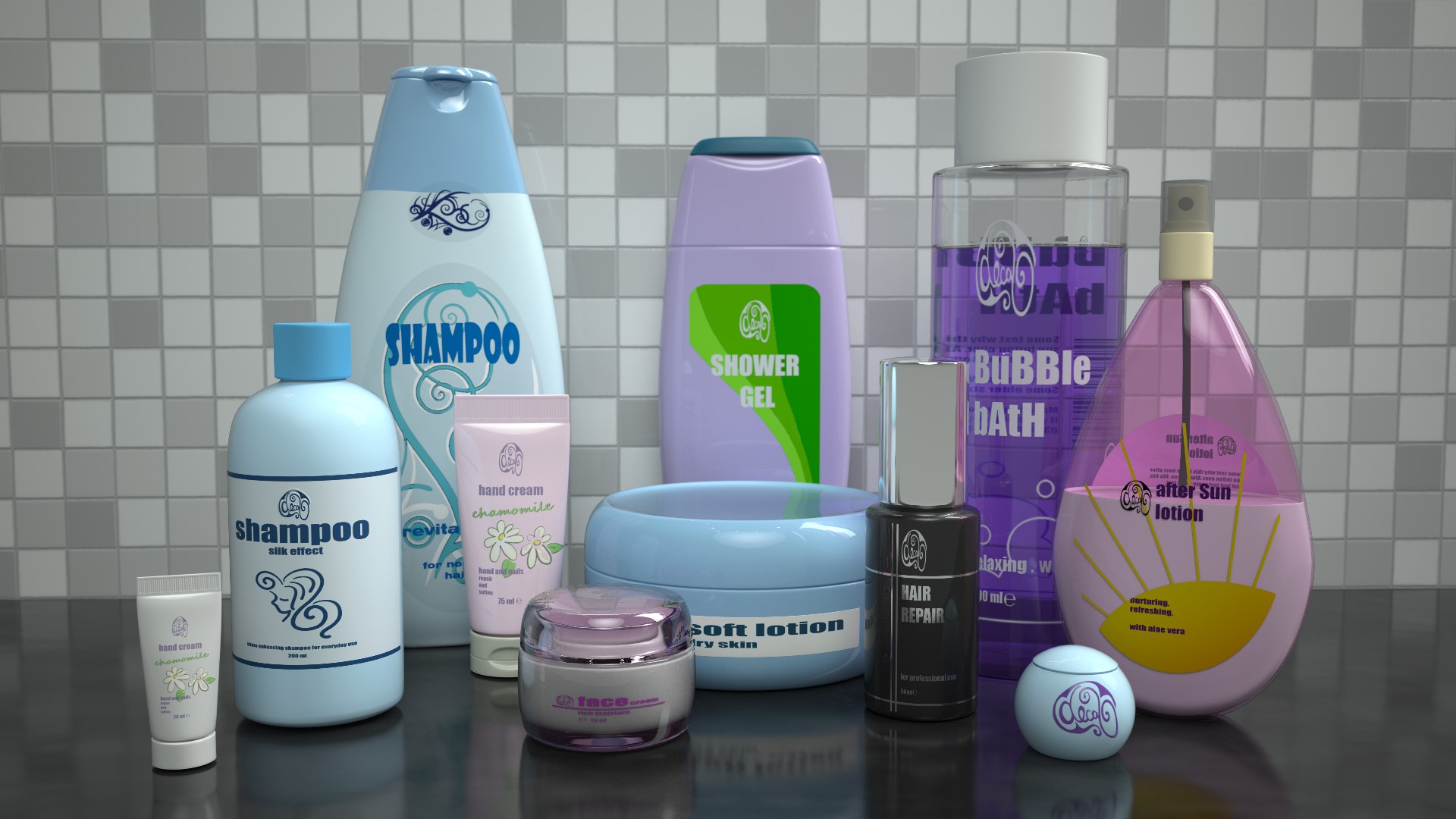}
    \caption{A variety of decals, projected onto cosmetic products. \label{Fig:Decals}}
\end{figure}

Upon intersecting an object with decals attached, the list of active decals is loaded and the topmost decal's material is chosen as active material.
From that point on we simply work with that material on the regular geometry.
The material then drives the decal's appearance and shape, and the MDL cutout property determines whether the currently processed decal is visible at the intersection point or not.
When the simulation would normally switch to continue ray tracing (e.g.\ when handling a cutout area), we skip this step, and instead continue to advance through the decal list, where reflection rays return to the previous decal and transmission rays proceed to the next.
The new decal then simply replaces the material and we continue with the computation while staying on the same geometry.
If eventually an end of the list is reached, an actual ray is cast to continue the global simulation.
Similarly, shadow rays for importance sampled light sources may pass through several decal layers.
In order to avoid duplicate cutout evaluations (light and reflection rays often pass the same layers), we cache them.

The placement of decals can be easily controlled by their position in the scene graph: if a decal is attached to a node, all of the child nodes are affected by the decal.
Priorities can be assigned to change the order, and the potentially different front or back side placements can be configured (e.g.\ to have different properties for decals placed on transmissive materials).
To further ease the placement, in particular across multiple objects with different UVW layouts, a projector function can be attached.
This projector function overrides the texture space for the decal material with procedurally generated UVWs.
Several kinds of projector functions are supported, like a planar projection for flat surfaces and a cylindrical projection that can be used to place labels onto bottles.
As an additional control for its spatial extent, a decal should also feature a clip box.
Each object that is potentially affected by decals then uses a bounding volume hierarchy over all clip boxes that intersect that object to quickly determine which decals have to be taken into account when handling a light path interaction.

The generality of Iray's decal implementation makes it easily possible to create advanced decal setups, for example a slightly transparent paper
label on a glass bottle which has a print on the front side and no print on the back side, or several semi-transparent plastic stickers on top of each other (Fig.~\ref{Fig:Decals}).
The downside of this concept is that with increasing complexity, e.g.\ with many semi-transparent decals and large overlap, the simulation cost increases accordingly.
In particular, if the decal clip box mechanism is not utilized at all, many cutout texture slots will be evaluated, too.

\section{Discussion and Limitations} \label{Sec:Discussion}

Some of Iray's limitations follow directly from the design decisions outlined in Sec.~\ref{Sec:Design}, others are due to tradeoffs
made while meeting those objectives, or are just a property of our current implementation without being inherent to the design decisions.
In this section, we discuss limitations from all of these categories.

\subsection{Metropolis Sampling}

While difficult scene setups can benefit from Metropolis sampling \cite{CKelemen}, it
unfortunately does not fit the wavefront state machine
execution model very well, as the number of parallel Markov
chains executed simultaneously is far from sufficient to take advantage of all available compute units:
To reduce runtime storage, only 32k parallel Markov chains can be kept around and simulated simultaneously in Iray.
In addition, Metropolis sampling actually benefits most from long chains and so it is better to have fewer long chains than
a larger number of short chains.
Each chain also adds significant startup bias by design, so again, increasing
the number of chains would amplify this problem, along with increasing storage (roughly 1~kB per chain).
Finally, the inherently sequential nature of Markov chain Monte Carlo also does not allow one to use path state
regeneration to increase processor utilization.

As an additional burden, Metropolis introduces the need for pseudo-random samples in the simulation and to store the state of the generator.
To reduce storage, the period of the pseudo-random number generator was limited by $2^{127}$.

\subsection{Workload}

Iray's state machine performs optimally when there is sufficient work available for path regeneration, thus keeping the 
kernel queues full and GPU utilization high. Therefore, it is best suited for batch rendering where each device renders full frames 
with many iterations. When rendering in interactive mode on many GPUs, the total work load per GPU drops and the benefits of 
state machine execution diminish. Similarly, the performance benefits are limited when rendering motion blur (see Sec.~\ref{Sec:MotionBlur}), because the number 
of samples per time step is usually quite low. Furthermore, the fixed time step motion blur method 
does not play well with Metropolis sampling, which requires long Markov chains to pay off. Using a 
motion aware acceleration data structure may overcome these issues at the cost of reduced ray tracing performance and potentially 
persistent motion vector approximation artifacts for both geometry and materials \cite{GrunschloB:2011:MEA:2018323.2018334}.

\subsection{Reordering}

Contrary to earlier work \cite{Hoberock:2009:SCD:1572769.1572797,journals:cgf:EisenacherNSB13,Wavefront},
reordering path states based on materials turned out to not help performance in the majority of scenes. As
Iray is quite heavy in path state access, trading coherent path state memory access for coherent material evaluation
does not pay off. An additional reason is found in the nature of complex materials: 
Due to the layering of multiple BSDFs and their often spatially varying parameter inputs, the ``random'' selection during the 
simulation can still result in severe code- and data-divergence within a warp, even though the same material is handled. On the 
other hand, significant portions of the material sampling and evaluation code are identical for most materials (see Sec.~\ref{Sec:MatEval})
and do not benefit from sorting either. Finally, Iray usually only generates up to one million paths in parallel on a single GPU.
This severely limits the amount of coherence that can be extracted by sorting, especially in common, material heavy scenes.
Similarly, experiments with sorting states by ray position and direction only showed very limited performance improvements in ray tracing.

\subsection{Heterogeneous Volumes}

Iray's current implementation only supports homogeneous volumes.
There is no fundamental obstacle in adding this capability to a renderer of the proposed architecture, and it is very likely that this feature will be added in a future version.
Still, it should be considered that the interaction of overlapping heterogeneous volumes (e.g.\ a flame in smoke) has to be specified.
The required user interaction is something that we would like to avoid, in order to not compromise the push-button usage.

\subsection{Firefly Handling}

In physically-based renderers, certain effects may suffer from very high variance due to
imperfect importance sampling, which may cause high frequency noise in the output. 
To eliminate this kind of noise, Iray can be configured to detect outliers in a user-driven or automatic fashion
(based on camera exposure settings) during the simulation process and clamp them to a reasonable range. 
While this approach may lead to a loss of energy in the final image, the bias introduced is negligible in the majority of scenes.
For predictive rendering the firefly filter needs to be turned off.

\subsection{No Out of Core Data}

To simplify data access and maximize performance, Iray is not required to support out of core scene data, i.e.\ each GPU holds a full copy of all scene data.
This may limit the size of supported scenes on low-end and older generation GPUs.

However, this limitation does not apply to the outputs of Iray, as framebuffers can be of arbitrary size
and there can be multiple outputs enabled at the same time (see Sec.~\ref{Sec:LPE}).

As current GPUs feature memory sizes of up to 24GB and the core also supports instancing of objects, scenes originating from the application domain of Iray so far also did not exceed this constraint.
For more complex scenes such as those seen in the visual effects industry, this is of course insufficient. Possibilities to overcome this limitation include unified virtual memory (UVM) or NVLINK.

\section{Conclusion}

The unique and careful design of the Iray light transport simulation and rendering system's architecture enables easy integration and use in new or existing products, in particular since
the consistent push-button approach does not lead to complex configuration interfaces, but greatly reduces training effort and enables obtaining predictable results even for novice users.
Features that follow directly from the design decisions (see Sec.~\ref{Sec:Design}) allow for practical, flexible, and efficient workflows.
As the high utilization of modern GPUs is retained when scaling across multiple devices and networks,
the performance required for high quality results and interactive rendering can be provided. 

Iray started to be shipped within 3ds Max in 2010,
is in active use by industry professionals today, and has led to the adoption of physically-based light transport simulation in many
industry solutions, among them
Catia Live Rendering and Solidworks Visualize (former Bunkspeed Shot) by Dassault Syst{\`e}mes,
Siemens NX,
Allegorithmic Substance Designer and Painter,
Lumiscaphe,
Sketchup via Bloom Unit, RealityServer and BIM IQ/Iray for Revit from migenius, 
DAZ Studio,
Iray for Maya,
Iray for 3ds Max,
Iray for CINEMA 4D,
Iray for Rhino, and 
as one rendering mode within mental ray, including the integration into 3ds Max and m4d Iray for CINEMA 4D.

While some challenges still lie ahead, we believe that the Iray light transport simulation and rendering system takes a unique place in today's and tomorrow's ecosystem of renderers.

\section*{Acknowledgements}

The authors would like to thank Michael R. Kaplan and Daniel Levesque as well as the product management, QA, development, and research teams at NVIDIA for their contributions to the Iray light transport simulation and rendering system.
In addition, the authors would like to thank Robert H\"odicke and Oliver Klehm for their valuable contributions to this paper.
Images and scenes were created and graciously provided by Jeff Patton, Paul Arden, and Delta Tracing, as well as Jan Jordan, Sandra Pappenguth, R\"udiger Raab, and Marlon Wolf of NVIDIA.

\clearpage
{\renewcommand{\markboth}[2]{}%
\printbibliography}

\clearpage
\newpage
\appendix

\section{Appendix} \label{Sec:Appendix}

\subsection{Anti-Aliasing} \label{Sec:AA}

The exact cause of what is usually called ``aliasing artifacts'' in rendered images is a common misconception in the field of computer graphics,
where the term ``aliasing'' does not always carry the exact same meaning as in a signal processing context.
We separate two different problems that arise when rendering and displaying images:
\begin{enumerate}
	\item Computing an integral over a complicated function (such as solving the rendering equation) over the area of each pixel and
	\item reconstructing a (mostly) continuous signal on a screen made up of (sub-)pixels.
\end{enumerate}

Estimating the integral was already discussed in Sec.~\ref{Sec:Techniques} and will not cause artifacts,
assuming that the integral can be solved in a numerically robust way and that the image is allowed to converge fully.
The result is effectively a box-filtered representation of the original signal, sampled at intervals corresponding to the filter kernel (pixel) size.
One is not able to reconstruct the original signal from this representation though, especially given the additional limitations of current display technology.
Samples are displayed directly on light emitters of a certain finite size, more or less corresponding to the size of the implicit box filter mentioned above.

The overall effect of this signal processing chain is somewhat similar in appearance to an effect that is well known from actual aliasing (i.e.\ undersampling in a regular pattern) during realtime rendering.
For example, aliasing often results in jagged edges on object boundaries, e.g.\ when only a single or very few samples per pixel completely determine the appearance of the whole pixel on screen, which amounts to using a simple box function for reconstructing an already undersampled high frequency signal.
As noted earlier, a very similar visual effect will appear in any image though, even if the integral is fully solved for each pixel, and is often called ``aliasing'' as well.
Especially in small areas of very high dynamic range, such as on the boundary of a very bright foreground object over a dark background, this effect becomes noticeable.
Pixels overlapping this boundary will exhibit brightness values corresponding to the proportion of foreground and background they cover.
Even with hypothetical future display technology that could directly display HDR values and make a tone-mapping step obsolete, any pixel that only slightly overlaps the bright part of the input signal will exhibit a brightness increase over its whole area on the display.

While the simulation itself cannot be made \textit{more correct} to solve this, applying a low-pass filter to the input signal can help to address this issue.
Instead of sharing samples in the image plane across overlapping supports of finite filter kernels,
Iray applies importance sampling on the filter kernel during ray generation.
This drastically simplifies the implementation in a highly parallel simulation environment, avoids problems at image boundaries, and can even result in noise reduction \cite{FilterImportanceSampling}.
Our implementation supports a variety of low-pass filters, as other filter types will introduce new artifacts for the same reasons as already discussed.
The guideline for our users is to use a Gaussian
filter covering nine pixels in its $3\sigma$ range, which is able
to resolve most common sources of ``aliasing'' due to its infinite support,
like edges of very bright light sources or reflections of these on highly specular materials.

\subsection{Quasi-Monte Carlo based Sequence Evaluation}  \label{Sec:Halton}

The faster convergence of using quasi-Monte Carlo methods \cite{Nie:92,NutshellQMC} over Monte Carlo methods \cite{Ermakov:75,Sobol:94} in computer graphics often comes at the
price of regular patterns in the early stages of computation. However, due to consistency, these artifacts are guaranteed to vanish.
Earlier versions of Iray were based upon the popular Sobol' sequence \cite{Sobol:67},
but suffered from such temporary patterns in certain scene setups, so we replaced it with the Halton sequence \cite{Hal:64} recently.

As the Halton sequence in its classical definition can be prone to even more temporary artifacts, especially when having to use
very large prime numbers as the base (like it is the case in light transport simulation that evaluates hundreds of dimensions before paths are cut off or absorbed), we use a rather simple, but effective scrambling \cite{Fau:2009} instead. This improves the image quality for practical sample counts
(e.g.\ hundreds to thousands of samples per pixel) dramatically.

Great care has to be taken when implementing the numerical evaluation of Halton based samples though. While the implementation can easily be made
precise when using base 2 (as it is equivalent to the binary radical inverse), any other base is usually implemented by employing floating
point arithmetic. 
This can lead to severe numerical issues, even when doing the computations in double precision.

Using full integer arithmetic instead, the implementation can be made numerically precise, but will then suffer from decreased performance, as the
iterative process triggers a large amount of divisions as the sample indices get larger during the simulation.
As Iray also enumerates samples over the full screen (without any tiling of samples) \cite{SampleEnum}, sequence indices can need up to 64 bits to be stored, which then need
to be evaluated in the Halton code.
By using strength reduction, an integer division can be replaced by an extended multiplication and some additional simple integer arithmetic
in a preprocess, which brings the overall performance back into the same ballpark as the optimized Sobol' sequence evaluation that Iray used before.

\subsection{Local Shading Normal Adaption}

\begin{figure}
	\centering
	\begin{overpic}[width=\linewidth]{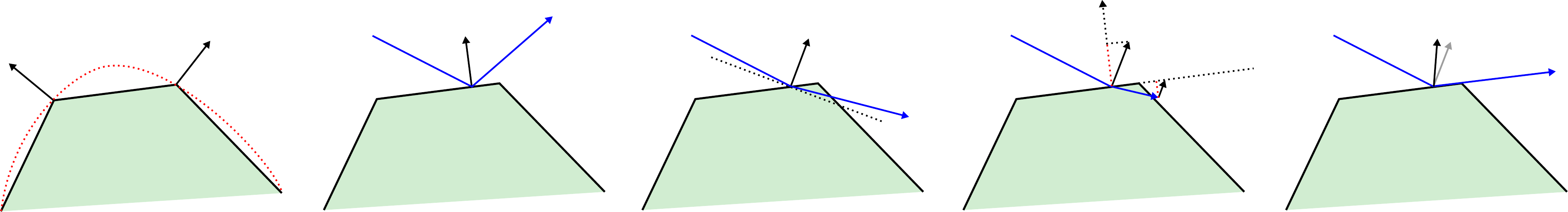}
		\newcommand{\s}{\scriptsize}
		\put(0.5,10){\s$N_{v1}$}\put(14,11){\s$N_{v2}$}
		\put(28,12){\s$N_g$}
		\put(51.8,11){\s$N_s$}
		\put(68.5,13.5){\s$N_g$}\put(72.5,11){\s$N_s$}\put(69.5,9){\s\color{red}$b$}\put(73,9){\s\color{red}$a$}\put(75,7){\s$\frac{a}{b}N_s$}
		\put(89,12){\s$N_s'$}\put(93,11){\s\color{gray}$N_s$}
	\end{overpic}
	\caption{Illustration of Shading Normal Adaptation.
		A smooth surface is approximated by triangles and vertex normals $N_{v1}$
		and $N_{v2}$. While an incoming ray reflected off the flat surface
		(geometric normal $N_g$) will never lead to self intersection, rays
		reflected off the virtual surface (interpolated shading normal $N_s$)
		will, depending on the surface approximation and the incoming ray.
		Replacing the shading normal with $N_s'$ will lead to no self intersection
		for perfect reflections. Note that in this 2D illustration the main
		benefit of keeping the original 3D orientation of incoming ray and shading normal
		intact is not immediately obvious. %
		\label{Fig:ClampNormalIllu}}
\end{figure}

\begin{figure}
	\centering
	\includegraphics[width=0.4915\linewidth]{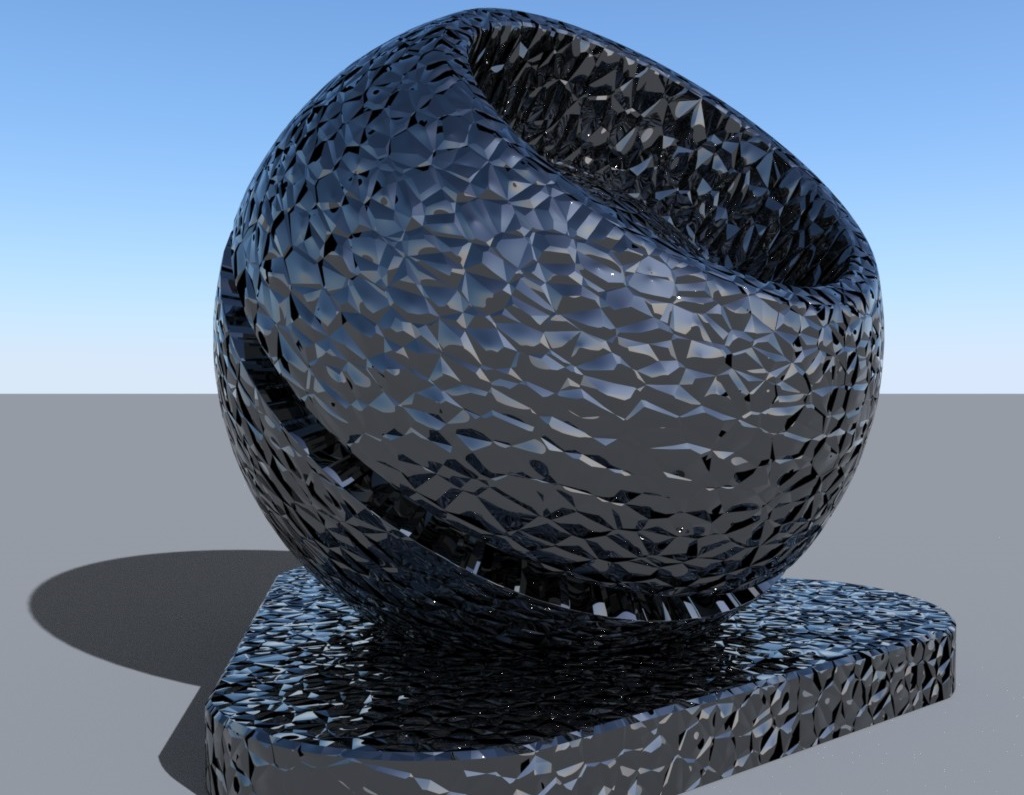} \hfill
	\includegraphics[width=0.4915\linewidth]{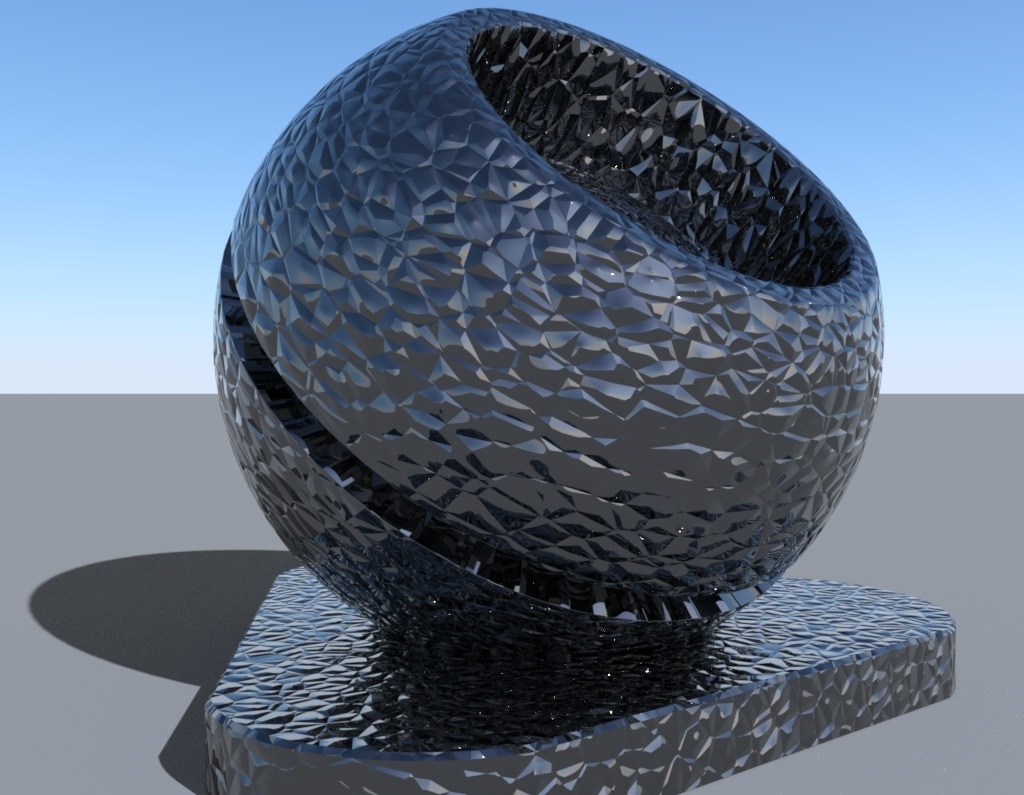}
	\caption{The effect of Shading Normal Adaptation on an object with smooth geometry and a normal map (Left: adaptation off, Right: adaptation on). Note the significant energy loss in the left picture (black spots towards the edges and on the plate). \label{Fig:ClampNormal}}
\end{figure}

Shading normals, and in particular strongly bumped normals,
are prone to lighting artifacts, especially on reflections, and can lead
to severe energy loss if they deviate a lot from the geometric normal.
The cause for that is that shading normals feign the presence of detailed geometry to the BSDF computation while the ray tracer actually only deals with flat surfaces.
Thus significant parts of the BSDF may yield no contribution as the associated rays cause self intersection with the flat geometry.

To mitigate such problems, Iray bends the main shading normal and the bumped normals on all layers, such that
the perfect reflection does not extend below the geometric tangent plane. This handles the corner case of perfectly specular materials well (e.g.\ avoids all self intersections in this case), but is also able to handle glossy effects nicely (e.g.\ avoids most self intersections for common BSDF models), and ensures that the incoming ray and shading normal agree on sidedness.
We achieve that by computing the perfect reflection vector (according to shading normal) and then, if that extends below surface, we ``pull it up'' towards the geometric normal such that it lies slightly above the tangent plane (see Fig.~\ref{Fig:ClampNormalIllu}).

With this new reflection vector a new shading normal can be computed (as the halfway vector to the incoming direction) that fulfills our requirement.
This solution provides a smooth result, is free of parameters, and also does not need to
rely on additional connectivity information of neighboring vertices or triangle
edges.

Note that while this scheme is view-direction-local (thus implicitly changing
the BSDF) and not physically plausible (which is a general problem with shading
normals) it performs very well in practice and, in particular, does not cause
any additional rendering artifacts.

\subsection{Driver Timeout Prevention and Recovery}

Most operating systems restart a graphics device driver if the runtime of a single GPU kernel exceeds a certain threshold, to prevent
a potential lockup of the user interface of the operating system.
In order to prevent such driver timeouts, Iray needs to limit the samples simulated per kernel. %
At the same time though, more samples per kernel allow for more parallelization on the GPU and thus better performance.
Iray balances these two competing goals using a timeout prevention and recovery mechanism.
Based on observed performance, the set of samples of long running kernels is split, aiming for about $1$ second execution time per kernel (or less if the GPU is also used for driving the user interface).
When the execution time of a running kernel approaches the threshold, the host communicates through mapped memory to the GPU kernel that a timeout is imminent.
The kernel checks this mapped memory regularly and aborts further computations if needed to prevent a forced driver restart.
Using a sample count buffer, each device keeps track of the sample count per pixel. 
If a OS timeout occurred, the sample count buffer is used to continue rendering in a subsequent kernel execution.

%
\end{document}